\documentclass[journal,draftcls,onecolumn,12pt,oneside]{IEEEtran}

\usepackage{cite}
\usepackage{float}
\usepackage{graphicx}
\usepackage{amsmath}
\usepackage{times}
\usepackage{graphicx}
\usepackage{latexsym}
\usepackage{graphicx}
\usepackage{bm}
\usepackage{amssymb}
\usepackage[center]{caption2}
\usepackage{stfloats}
\usepackage{cases}
\usepackage{array}
\usepackage{setspace}
\usepackage{fancyhdr}
\usepackage{color}
\usepackage{subfigure}
\usepackage{multirow}
\usepackage{amsmath}
\usepackage{cases}
\usepackage[noend]{algpseudocode}
\usepackage{algorithm}
\usepackage{algorithmicx}
\usepackage{diagbox}
\usepackage{algpseudocode}
\usepackage{epstopdf}
\usepackage{varwidth}
\usepackage{pifont}
\usepackage{mathrsfs}
\usepackage{booktabs} 

\usepackage{verbatim}

\usepackage[colorlinks,urlcolor=blue,linkcolor=black]{hyperref}

\usepackage{url}

\newcommand{\norm}[1]{\left\|#1\right\|}

\begin{document}

\title{Spatially Sparse Precoding in Wideband Hybrid Terahertz Massive MIMO Systems}
\author{\IEEEauthorblockN{Jiabao Gao, Caijun Zhong,~\IEEEmembership{Senior Member,~IEEE,} Geoffrey Ye Li,~\IEEEmembership{Fellow,~IEEE}, Joseph B. Soriaga, and Arash Behboodi,~\IEEEmembership{Member,~IEEE}
		\thanks{J. Gao and C. Zhong are with the College of Information Science and Electronic Engineering, Zhejiang University, Hangzhou 310007, China (e-mail: gao\_jiabao@zju.edu.cn; caijunzhong@zju.edu.cn).}
		\thanks{G. Y. Li is with the Department of Electrical and Electronic Engineering, Imperial College London, London SW7 2BU, U.K. (e-mail: Geoffrey.Li@imperial.ac.uk).}
		\thanks{J. B. Soriaga is with Qualcomm Technologies, Inc., 5775 Morehouse Dr, San Diego CA 92122 (e-mail: jsoriaga@qti.qualcomm.com).}
		\thanks{A. Behboodi is with Qualcomm Technologies Netherlands B.V., Science Park 400, 1098 XH Amsterdam (e-mail: behboodi@qti.qualcomm.com).}
}}

\maketitle

\begin{abstract} 
	In terahertz (THz) massive multiple-input multiple-output (MIMO) systems, the combination of huge bandwidth and massive antennas results in severe beam split, thus making the conventional phase-shifter based hybrid precoding architecture ineffective. With the incorporation of true-time-delay (TTD) lines in the hardware implementation of the analog precoders, delay-phase precoding (DPP) emerges as a promising architecture to effectively overcome beam split. However, existing DPP approaches suffer from poor performance, high complexity, and weak robustness in practical THz channels. In this paper, we propose a novel DPP approach in wideband THz massive MIMO systems. First, the optimization problem is converted into a compressive sensing (CS) form, which can be solved by the extended spatially sparse precoding (SSP) algorithm. To compensate for beam split, \emph{frequency-dependent} measurement matrices are introduced, which can be approximately realized by feasible phase and delay codebooks. Then, several efficient atom selection techniques are developed to further reduce the complexity of extended SSP. In simulation, the proposed DPP approach achieves superior performance, complexity, and robustness by using it alone or in combination with existing DPP approaches.
\end{abstract}


\newpage

\section{Introduction}
In future 6G wireless communication systems, it is foreseeable that ultra-high data rates will be required to support many cutting-edge applications, such as virtual reality, augmented reality, and digital twins\cite{towards_6G}. With abundant bandwidth resources, terahertz (THz) is a promising frequency band to meet the sharply increased data rate requirement\cite{THz1,THz2}. Specifically, the bandwidth provided by the THz band could reach up to several tens of GHz, which is much larger than that in the millimeter-wave (mmWave) band used in 5G\cite{THz1}. However, higher carrier frequencies make the THz signals suffer from more severe attenuation than signals in lower frequency bands\cite{massive_mimo1}. To alleviate this issue, THz is usually combined with massive multiple-input multiple-output (MIMO), where directional beams with high array gains can be generated by massive antennas at transceivers\cite{massive_mimo1,massive_mimo2,massive_mimo3}. Furthermore, to relieve the manufacturing cost and power consumption, it is a common practice to adopt the hybrid precoding architecture in THz massive MIMO systems, where precoding is decomposed into a high-dimensional analog precoder and a low-dimensional digital precoder\cite{hbf1,hbf2}. Thanks to the sparsity of THz channels in the angular domain, satisfactory performance can still be achieved with only a few radio-frequency (RF) chains using hybrid precoding. Next, we will briefly review prior works about hybrid precoding in THz massive MIMO systems. Since mmWave channels and THz channels have many similar properties, some works based on the mmWave model are included as well.

\subsection{Prior Works}
In most prior works, the analog precoder is implemented by phase shifters. Since the optimal solutions are hard to obtain in the presence of the non-convex unit modulus constraint of phase shifts, the approximate orthogonal properties of the optimal analog and digital precoders are exploited in \cite{assumption_simplify} to simplify the optimization process. In the well-known \emph{spatially sparse precoding} (SSP) algorithm\cite{ssp}, the sum rate maximization problem is first converted to an approximately equivalent while much simpler matrix decomposition problem, where the distance between the fully-digital precoder and the hybrid precoder is minimized. Then, by exploiting the structures of optimal precoders, the optimization problem is further converted to a typical sparse recovery problem that can be readily solved by various compressive sensing (CS) algorithms. In \cite{ssp_low_complexity}, the complexity of SSP is reduced by using orthogonal codebooks and beamspace singular value decomposition (SVD). To achieve better performance without constraining the form of the analog precoder, an alternative minimization (AM) algorithm is proposed in \cite{AM_based} to optimize the analog and digital precoders alternately. In \cite{DL_HBF1,DL_HBF2}, deep neural networks are used to predict hybrid precoders based on channel state information (CSI). Without the need of estimating the high-dimensional full CSI, codebook based approaches are appealing in practice, where efficient codebooks and beam sweeping procedures are developed to reduce the overhead\cite{codebook_based1,codebook_based2}. 

Although the above phase-shifter-based hybrid precoding architecture works well in most narrowband systems\cite{hbf2}, it will suffer from severe performance degradation in wideband systems, especially when the number of antennas and the fractional bandwidth (ratio between bandwidth and central frequency) are large, e.g., THz massive MIMO. The reason is the existence of the \textit{beam split} effect\footnote{Notice that the term beam squint is used more often in mmWave systems. Since beam split can be seen as the more severe version of beam squint, we will just use the term beam split in both mmWave and THz systems for neatness in this paper.}, which is proportional to both the number of antennas and the fractional bandwidth\cite{beam_split_effect1,beam_split_effect2,gff_heuristic}. With beam split, there is a natural contradiction, such that the same physical channel path will have different equivalent angles at different subcarriers and make the optimal fully-digital precoders \textit{frequency-dependent} while the analog precoder implemented by phase shifters is shared by all subcarriers and \textit{frequency-independent}\cite{gff_heuristic}. 

To minimize the impact of beam split, several methods have been proposed from the perspective of \textit{algorithm redesign}\cite{beam_split_no_TTD_1,beam_split_no_TTD_2,beam_split_no_TTD_3,beam_split_no_TTD_4,gff_heuristic}. In \cite{gff_heuristic}, a beam broadening approach has been proposed where wider beams are constructed through subarray coordination to cover the split directions at all subcarriers. Since the AM algorithm works with general channels including those with beam split, it is used in \cite{beam_split_no_TTD_1} to solve the wideband hybrid precoding problem. More generally, AM is used in \cite{beam_split_no_TTD_2} to develop mmWave and THz hybrid precoding algorithms for various analog architectures, such as the quantized phase shifters and the use of switches for dynamic connections. In \cite{beam_split_no_TTD_3,beam_split_no_TTD_4}, codebooks are revised to maximize the minimum array gain and the total array gain achieved across all subcarriers, respectively. Nevertheless, these methods still use a frequency-independent analog precoder, therefore will lose most of the effectiveness when beam split is relatively severe.

Since beam split essentially comes from the delay of electromagnetic wave travelling through the antenna array\cite{beam_split_effect1}, a more direct and thorough solution is to compensate for the delay from the perspective of \textit{hardware redesign}. In the emerging delay-phase precoding (DPP) architecture, true-time-delay (TTD) lines are incorporated apart from phase shifters to implement analog precoders. Since the phase shift of a TTD line is proportional to both its delay parameter and the subcarrier frequency, \textit{frequency-dependent} analog precoders can be realized. Assuming the path channel model, heuristic DPP algorithms that exploit the angles of channel paths sorted by their path gains are proposed in \cite{gff_heuristic,DPP}, which can achieve near-optimal performance with low complexity. To support cluster channels, the AM algorithm is used in \cite{Alt_Opt} to optimize the time delays, phase shifts, and digital precoders alternately. To improve energy efficiency, fixed TTD lines are used along with a switch network in \cite{fixed_TTD}. In \cite{subarray_codebook}, multi-resolution time-delay codebooks are designed through subarray coordination for efficient beam sweeping, while the level of beam split is controlled by delay-phase configuration to accelerate beam tracking in \cite{wideband_beam_tracking}. In \cite{DPP_practical_channel}, the effectiveness of the DPP architecture is validated by practical 6G THz massive MIMO channels.

Among existing DPP algorithms, it is hard to find one that satisfies both the strict requirements for performance and complexity in 6G communication systems with practical THz channels. Specifically, the performance of heuristic DPP algorithms will degrade in cluster channels while iterative optimization based DPP algorithms usually suffer from high complexity. The SSP algorithm originally designed for narrowband systems has many attractive characteristics thanks to the exploitation of precoder structures\cite{ssp}. Nevertheless, its extension to wideband DPP with beam split considered is not investigated yet.

\subsection{Contributions}
In this paper, we first propose a novel extended SSP algorithm to solve the DPP problem in wideband THz massive MIMO systems and then propose several tailored techniques to further reduce the algorithm's complexity. The main contributions are as follows:
\begin{itemize}
\item Following the key idea of SSP, we first formulate the wideband DPP problem as a matrix decomposition problem to approximate fully-digital precoders with hybrid precoders at all subcarriers. Then, the optimization problem is further formulated as a multiple-measurement-vector (MMV) CS problem by exploiting the structures of optimal precoders. Eventually, the simultaneous orthogonal matching pursuit (SOMP) algorithm is used to solve the MMV CS problem. 
\item To recover the common sparsity structure destroyed by beam split and sustain the effectiveness of SOMP, we first design ideal \emph{frequency-dependent} measurement matrices to align the sparse supports of projections at all subcarriers. Then, feasible phase and delay codebooks that can be implemented by hardware including phase shifters and TTD lines are optimized to approximately realize the ideal matrices.
\item By exploiting the properties of projections in SOMP in the angular-frequency domain, we further develop several efficient atom selection techniques to reduce the complexity of the proposed DPP algorithm. Specifically, the number of iterations, atoms, and subcarriers involved in atom selection is reduced significantly without sacrificing much performance. Simulation results illustrate that the proposed approach can be applied alone to achieve satisfactory performance with very low complexity, or serve as a cheap yet effective initializer to improve the convergence performance and speed of the existing high-performance iterative optimization based DPP algorithm. 
\end{itemize}

The rest of this paper is organized as follows. In Section II, the hybrid massive MIMO system, the wideband THz channel model, and the DPP optimization problem are introduced. Then, the extended SSP algorithm is elaborated in Section III, whose complexity is further reduced using the efficient atom selection techniques presented in Section IV. In Section V, the superiority of the proposed approach is validated through numerical results, and the paper is eventually concluded in Section VI.

\emph{Notations:} Italic, bold-face lower-case and bold-face upper-case letters denote scalar, vector, and matrix, respectively. $(\cdot)^{*}$, $(\cdot)^{T}$, $(\cdot)^{H}$, $(\cdot)^{\dagger}$, $\mathbb{E}(\cdot)$, $\odot$, and $\otimes$ denote conjugate, transpose, conjugate transpose, Moore-Penrose inverse, expectation, Hadamard product, and Kronecker product, respectively. $\bm{A}^{-\frac{1}{2}}$ denotes the matrix that satisfies $\bm{A}^{-\frac{1}{2}}\bm{A}^{-\frac{1}{2}}=\bm{A}^{-1}$. $\norm{\bm{A}}_F$ denotes the Frobenius norm of matrix $\bm{A}$ and $\left\| \bm{a} \right\|_p$ denotes the $p$-norm of vector $\bm{a}$. $|a|$ denotes the modulus of complex scalar $a$ while $|\bm{A}|$ denotes the determinant of matrix $\bm{A}$. $\bm{I}_N$ denotes the $N\times N$ identity matrix while $\bm{1}_{N}$ and $\bm{0}_{N}$ denote the $N\times 1$ all-one and all-zero column vectors. $\text{diag}(\cdot)$ can denote the diagonalization operation or the inverse diagonalization operation depending on specific situations. $a // b$ denotes the operation of taking the quotient of the dividend $a$ and the divisor $b$. Inheriting the Matlab style of matrix slicing, we use $A(i,j)$, $\bm{A}(i,:)$, $\bm{A}(:,j)$, and $\bm{A}(:,i:j)$ to denote the element at the $i$-th row and $j$-th column, the elements at the $i$-th row, the elements at the $j$-th column, and the elements from the $i$-th column to the $j$-th column of matrix $\bm{A}$, respectively. $[\cdot|\cdot]$ can denote the horizontal or vertical concatenation of matrices or vectors depending on specific situations. $a:b:c$ denotes the arithmetic sequence vector that starts from $a$ and ends at $c$ with common difference $b$. ${\mathbb C^{x \times y}}$ denote the ${x \times y}$ complex space. $\mathcal{CN}(\mu,\sigma^2)$ denotes a circularly symmetric complex Gaussian (CSCG) random variable with mean $\mu$ and variance $\sigma^2$.

\section{System model and problem formulation}
In this section, we first introduce the hybrid massive MIMO system under the DPP architecture and the practical wideband THz cluster channel model along with beam split. Then, the DPP optimization problem and its conversion to the matrix decomposition form are elaborated. 

\subsection{System and Channel Models}
\begin{figure}[!htb]
	\centering
	\includegraphics[width=1\textwidth]{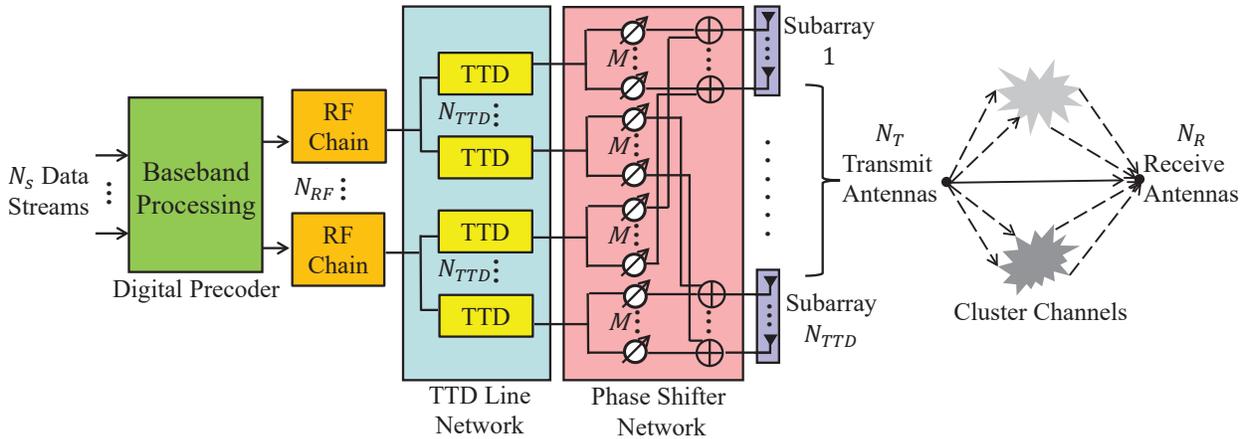}
	\caption{Hybrid massive MIMO system under the DPP architecture with cluster channels.}
	\label{system}
\end{figure}

We consider the hybrid massive MIMO system illustrated in Fig. \ref{system}, where a base station (BS) with an $N_T$-antenna uniform linear array (ULA) and $N_{RF}$ RF chains serves a user with an $N_R$-antenna ULA. We first consider the single-user scenario while the extension to the multi-user scenario will be investigated in our simulation. The number of data streams is $N_s$. Usually, we have $N_s=N_R\le N_{RF}\ll N_T$ in practice\cite{DPP}. To combat the frequency selectivity of channels, OFDM is adopted where the total system bandwidth $f_s$ is evenly divided by $K$ subcarriers.

According to measurement results, THz channels usually have clustered characteristics\cite{cluster_channel}. Considering half-wavelength antenna spacing, we first define the $N$-dimensional ULA response vector as
\begin{equation}
	\bm{a}_N(\cdot)\triangleq[1,e^{-j\pi(\cdot)},\cdots,e^{-j\pi(N-1)(\cdot)}]^T/\sqrt{N},
\end{equation}
then the $N_R \times N_T$ channel matrix at the $k$-th subcarrier from the BS to the user can be expressed as
\begin{equation}
	\bm{H}_k=\sqrt{\frac{N_TN_R}{N_cN_p}}\sum_{i=1}^{N_c}\sum_{j=1}^{N_p}\alpha_{i,j}e^{-j2\pi f_k\tau_{i,j}}\bm{a}_{N_R}(\psi^R_{i,j,k})\bm{a}^T_{N_T}(\psi^T_{i,j,k}),
	\label{channel_model}
\end{equation}
where $N_c$ and $N_p$ denote the number of clusters and the number of subpaths in a cluster, respectively, $f_k=f_c+(k-1-\frac{K-1}{2})\eta$ is the frequency of the $k$-th subcarrier with $f_c$ and $\eta=\frac{f_s}{K}$ denoting the central frequency and the subcarrier frequency spacing, respectively. Besides, $\alpha_{i,j}$, $\tau_{i,j}$, $\psi^T_{i,j,k},\psi^R_{i,j,k}$ denote the complex gain, the delay, the equivalent angle of departure (AoD) at the BS, and the equivalent angle of arrival (AoA) at the user of the $j$-th subpath in the $i$-th cluster, respectively. Notice that, the equivalent path angles are dependent on subcarrier index $k$ due to beam split. Specifically, we have $\psi_{i,j,k}=\frac{f_k}{f_c}\mathrm{sin}(\theta_{i,j})$ at both transceiver sides\footnote{Please see \cite{beam_split_effect1} for detailed derivations of the frequency-dependent equivalent angles caused by beam split.}, where $\mathrm{sin}(\theta_{i,j})$ denotes the actual physical path angle. In the $i$-th cluster, we have $\theta^T_{i,j}=\bar{\theta}^T_i+\triangle\theta^T_j,\theta^R_{i,j}=\bar{\theta}^R_i+\triangle\theta^R_j,\tau_{i,j}=\bar{\tau}_i+\triangle\tau_j,\forall j$, where $\bar{\theta}^T_i,\bar{\theta}^R_i,\bar{\tau}_i$ denote the mean AoD, mean AoA, and mean delay, respectively, while $\triangle\theta^T_j,\triangle\theta^R_j,\triangle\tau_j$ follow the Laplacian distribution\cite{cluster_channel} with zero mean and standard deviation being $\sigma_{\theta^T_i},\sigma_{\theta^R_i},\sigma_{\tau_i}$, respectively. The standard deviation here is also known as the angular or delay spread. Without loss of generality, we assume $\alpha_{i,j}\in \mathcal{CN}(0,1),\forall i,j$, $\bar{\theta}^T_i,\bar{\theta}^R_i\sim \mathcal{U}[0,2\pi],\bar{\tau}_i\sim \mathcal{U}[0,\tau_{max}],\forall i$, where $\tau_{max}$ denotes the maximum delay. 

To effectively compensate for beam split, the DPP architecture is adopted. As illustrated in Fig. \ref{system}, a small-scale TTD line network is inserted between RF chains and the large-scale phase shifter network. To reduce the hardware cost and energy consumption, each RF chain only connects to $N_{TTD}$ TTD lines ($N_{TTD}\ll N_T$), and each TTD line further connects to a subarray consisting of $M$ antennas through $M$ phase shifters. RF chains and antennas are still fully-connected through the intermediate TTD line network such that $N_T=MN_{TTD}$. So, the $N_R\times 1$ user's received signal at the $k$-th subcarrier can be expressed as
\begin{equation}
\bm{y}_k=\sqrt{\rho}\bm{H}_k\bm{A}_k\bm{D}_k\bm{x}_k + \bm{n}_k,
\end{equation}
where $\bm{x}_k$ denotes the $N_s\times 1$ transmit signal at the BS at the $k$-th subcarrier satisfying $\mathbb{E}\{\bm{x}_k\bm{x}^H_k\}=\frac{1}{N_s}\bm{I}_{N_s}$, and $\rho$ denotes the transmit power at the $k$-th subcarrier\footnote{Although it is totally feasible to allocate power among subcarriers, here we consider equal power allocation among subcarriers as in \cite{ssp,Alt_Opt}.}. The analog and digital precoders at the $k$-th subcarrier are denoted by $\bm{A}_k\in\mathbb C^{N_T\times N_{RF}}$ and $\bm{D}_k\in\mathbb C^{N_{RF}\times N_s}$, respectively. The additive white Gaussian noise (AWGN) at the $k$-th subcarrier is denoted by $\bm{n}_k$, whose elements follow $\mathcal{CN}(0,\sigma_n^2)$ with $\sigma_n^2$ denoting the noise variance. The signal-to-noise ratio (SNR) is defined as $\rho/\sigma_n^2$.

The main difference between the DPP architecture and the conventional phase-shifter-based hybrid precoding architecture is that the analog precoder at each subcarrier is \emph{frequency-dependent}. Specifically, the analog precoder at the $k$-th subcarrier can be expressed as
\begin{equation}
\bm{A}_k=\bm{A}\odot(\bm{T}_k\otimes \bm{1}_M),
\label{feasible_matrices}
\end{equation}
where $\bm{A}$ denotes the common phase matrix introduced by the phase shifter network, and the $N_{TTD}\times N_{RF}$ matrix $\bm{T}_k$ denotes the \emph{frequency-dependent} phases introduced by the TTD line network at the $k$-th subcarrier. The Kronecker product of $\bm{T}_k$ and $\bm{1}_M$ comes from the one-to-$M$ connection mode between TTD lines and antennas at the BS. Denoting the delay matrix of the TTD line network as $\bm{T}$, a certain TTD line's phase at a certain subcarrier is proportional to both the delay parameter and the subcarrier frequency and can be expressed as
\begin{equation}
T_k(n_{TTD},n_{RF})=e^{-j2\pi f_kT(n_{TTD},n_{RF})},\forall n_{TTD},n_{RF},k.
\label{TTD_phase}
\end{equation}

\subsection{Problem Formulation}
We first assume that perfect downlink CSI is available at the BS\footnote{The downlink CSI can be obtained from the uplink CSI based on channel reciprocity in time division duplex (TDD) systems, or be estimated by the user and fed back to the BS in frequency division duplex (FDD) systems. Various existing algorithms can be used to estimate wideband THz channels with beam split properly handled\cite{beam_split_effect2,beam_split_ce1,beam_split_ce2,beam_split_ce3}.} and the impact of imperfect CSI will be investigated in simulation. Based on the downlink CSI, analog and digital precoders need to be designed to maximize the sum rate of all subcarriers, which can be expressed as 
\begin{equation}
R = \sum_{k=1}^{K}\mathrm{log}_2\left(\left|\bm{I}_{N_R}+\frac{\rho}{N_s\sigma_n^2}\bm{H}_k\bm{A}_k\bm{D}_k\bm{D}^H_k\bm{A}^H_k\bm{H}^H_k\right|\right).
\label{sum_rate}
\end{equation}

In the presence of constraints caused by the hardware used for analog precoding implementation, directly maximizing (\ref{sum_rate}) is difficult due to its complicated form. Therefore, we turn to solve the following matrix decomposition problem to approximate the fully-digital precoders with the hybrid precoders at all subcarriers, which is proved to be approximately equivalent to sum rate maximization while much simpler to solve\cite{ssp}: 
\begin{alignat}{2}
	\mathcal{P}1:\quad \mathop{\mathrm{min}}\limits_{\bm{A},\bm{T},\bm{D}_k}  &  \quad \sum_{k=1}^{K}\norm{\bm{F}_k-\bm{A}_k\bm{D}_k}_F^2, \\
	\mbox{s.t.} &\quad|A(n_{T},n_{RF})|=1, \forall n_{T},n_{RF}, \\
	&\quad0\ll T(n_{TTD},n_{RF}) \ll t_{max},\forall n_{TTD},n_{RF},  \\
	&\quad\norm{\bm{A}_k\bm{D}_k}_F^2=N_s,\forall k.
\label{P1}
\end{alignat}

In $\mathcal{P}1$, (8) is the unit modulus constraint of phase shifters' phase shifts, $t_{max}$ in (9) denotes the maximal delay of TTD lines, and (10) is to meet the transmit power constraint. To obtain the $N_T\times N_s$ fully-digital precoder at the $k$-th subcarrier, $\bm{F}_k$, SVD is first performed on the subchannel, i.e., $\bm{H}_k=\bm{U}_k\bm{\Sigma}_k\bm{V}^H_k$. Then, $\bm{F}_k$ is determined as $\bm{F}_k=\bm{V}_k(:,1:N_s)\text{diag}(\bm{p}_k)$, where the $N_s\times 1$ power allocation vector, $\bm{p}_k$, satisfying $\norm{\bm{p}_k}_2^2=N_s$, can be obtained by the well-known water-filling algorithm. For neatness, detailed mathematical procedures of water-filling based power allocation are omitted in this paper.

\section{Extended SSP Based Wideband DPP}
In this section, we first briefly introduce the principles of the conventional SSP algorithm designed for narrowband phase-shifter-based hybrid precoding. Then, we extend it properly to solve the considered wideband DPP problem.

\subsection{SSP Based Narrowband Hybrid Precoding}
Considering narrowband systems with only phase shifters, $\mathcal{P}1$ is simplified to the following optimization problem with $N_{TTD}=0$ and the subscript $k$ discarded:
\begin{alignat}{2}
	\mathcal{P}2:\quad \mathop{\mathrm{min}}\limits_{\bm{A},\bm{D}}  &  \quad \norm{\bm{F}-\bm{A}\bm{D}}_F^2, \\
	\mbox{s.t.} &\quad|A(n_{T},n_{RF})|=1, \forall n_{T},n_{RF}, \\
	&\quad\norm{\bm{A}\bm{D}}_F^2=N_s.
	\label{P2}
\end{alignat}
	
As one of the most classical algorithms to solve $\mathcal{P}2$, the key idea of SSP is that the fully-digital precoder can be approximated by the linear combinations of several array response vectors when the channel is sparse in the angular domain\cite{ssp}, e.g., in the mmWave or THz band. Based on such precoder structures, $\mathcal{P}2$ can be converted to the following CS form:
\begin{alignat}{2}
	\mathcal{P}3:\quad \mathop{\mathrm{min}}\limits_{\widetilde{\bm{D}}}  &  \quad \norm{\bm{F}-\widetilde{\bm{A}}\widetilde{\bm{D}}}_F^2, \\
	\mbox{s.t.} & \quad \norm{\text{diag}(\widetilde{\bm{D}}\widetilde{\bm{D}}^H)}_0=N_{RF}, \\
	&\quad\norm{\widetilde{\bm{A}}\widetilde{\bm{D}}}_F^2=N_s,
	\label{P3}
\end{alignat}
where the $N_T\times G$ measurement matrix consisting of $G$ atoms (e.g., columns) in the form of ULA response vectors is defined as
\begin{equation}
\widetilde{\bm{A}}\triangleq [\bm{a}_{N_T}(\phi^1),\bm{a}_{N_T}(\phi^2),\cdots,\bm{a}_{N_T}(\phi^G)],
\label{matrix_narrowband}
\end{equation}
with the corresponding angles evenly dividing the entire angular space such that 
\begin{equation}
	\phi^g=-1+(2g-1)/G,\forall g.
\end{equation}

Now, let us describe $\mathcal{P}3$ using CS terms. The columns of $\bm{F}$ are \emph{measurements}, $\widetilde{\bm{A}}$ is the common \emph{measurement matrix}, and the columns of the $G \times N_s$ auxiliary matrix $\widetilde{\bm{D}}$ are \emph{sparse vectors} corresponding to $N_s$ streams that have common sparse supports. The number of non-zero rows in $\widetilde{\bm{D}}$ equals the number of RF chains, which means that the analog precoding vector of each RF chain has to be one of the atoms of the measurement matrix. As a typical MMV-CS problem, $\mathcal{P}3$ can be readily solved by the slightly modified SOMP algorithm demonstrated in {\bfseries Algorithm 1}. In each of the $N_{RF}$ iterations, the atom with maximal summed projections of streams on the residual is selected to update the analog precoder while the digital precoder is updated by least square (LS) based on the temporal analog precoder and the fully-digital precoder. At the end, the power of the digital precoder is normalized to meet the transmit power constraint. 
\begin{algorithm}[!htb]
	\caption{SSP Based Narrowband Hybrid Precoding\cite{ssp}}
	\textbf{Input}: $\widetilde{\bm{A}},\bm{F},N_{RF};$ 
	\begin{algorithmic}[1]
		\State Initialize $\bm{A}^{(0)}$ as an empty matrix, initialize $\bm{F}_{\text{res}}^{(0)}=\bm{F};$  
		\State \textbf{for} $i=1:1:N_{RF}$ \textbf{do}  
		\State \quad $\bm{\Psi}^{(i)}=\widetilde{\bm{A}}^H\bm{F}_{\text{res}}^{(i-1)};$  
		\State \quad $\bm{\psi}^{(i)}=\text{diag}(\bm{\Psi}^{(i)}{\bm{\Psi}^{(i)}}^H);$
		\State \quad $g^{(i)}=\mathop{\mathrm{argmax}}\limits_{j}\psi^{(i)}(j);$
		\State \quad $\bm{A}^{(i)}=[\bm{A}^{(i-1)}|\widetilde{\bm{A}}(:,g^{(i)})];$
		\State \quad $\bm{D}^{(i)}={\bm{A}^{(i)}}^{\dagger}\bm{F};$
		\State \quad $\bm{F}_{\text{res}}^{(i)}=\frac{\bm{F}-\bm{A}^{(i)}\bm{D}^{(i)}}{\norm{\bm{F}-\bm{A}^{(i)}\bm{D}^{(i)}}_F};$
		\State \textbf{end for} 
	\State $\bm{A}=\bm{A}^{(N_{RF})};$
	\State $\bm{D}=\sqrt{N_s}\frac{\bm{D}^{(N_{RF})}}{\norm{\bm{A}^{(N_{RF})}\bm{D}^{(N_{RF})}}_F};$
	\end{algorithmic}
	\textbf{Output}: $\bm{A},\bm{D};$
	\label{algorithm_ssp}
\end{algorithm}

As indicated by \cite{DPP,Alt_Opt}, directly applying SSP to wideband DPP will result in poor performance, which is not surprising because beam split and the delay optimization of TTD lines are not considered in the original designs. Next, we will introduce the proposed extension of SSP from narrowband to the considered wideband DPP scenario.

\subsection{Frequency-Dependent Measurement Matrices}
\label{matrix_design}
Without beam split and TTD lines, the extension of SSP from narrowband to wideband would be straightforward. Since the equivalent path angles are the same at all subcarriers, the sparse supports of projections would be aligned along the subcarrier dimension. However, with beam split, such common sparsity structure among subcarriers no longer holds if a \emph{frequency-independent} measurement matrix is still used. As a result, target atoms that have good performance at all subcarriers cannot be accurately captured by simple averaging. Besides, using same atoms as analog precoders of all subcarriers cannot exploit TTD lines' ability of providing \emph{frequency-dependent} phases. To address this issue, we use \emph{frequency-dependent} measurement matrices\footnote{Notice that such an idea is previously used in \cite{beam_split_effect2} for wideband THz channel estimation. Thanks to the similar CS formulation, it can be naturally borrowed for the wideband THz precoding problem considered in this paper.}, where the measurement matrix used at the $k$-th subcarrier is defined as
\begin{equation} 
\widetilde{\bm{A}_{k}^{\text{ideal}}}\triangleq[\bm{a}_{N_T}(\phi_k^1),\bm{a}_{N_T}(\phi_k^2),\cdots,\bm{a}_{N_T}(\phi_k^{G})],
\label{ideal_matrices}
\end{equation}
with \emph{frequency-dependent} angles corresponding to atoms being
\begin{equation}
\phi_k^g=\frac{f_k}{f_c}\phi^g,\forall g.
\label{grids}
\end{equation}

\begin{figure}[htb!] 
	\centering 
	\vspace{-0.35cm} 
	\subfigtopskip=2pt 
	\subfigbottomskip=1pt 
	\subfigcapskip=-3pt 
	\subfigure[With the frequency-independent $\widetilde{\bm{A}}$.]{
		\label{impact_of_snr}
		\includegraphics[width=0.48\textwidth]{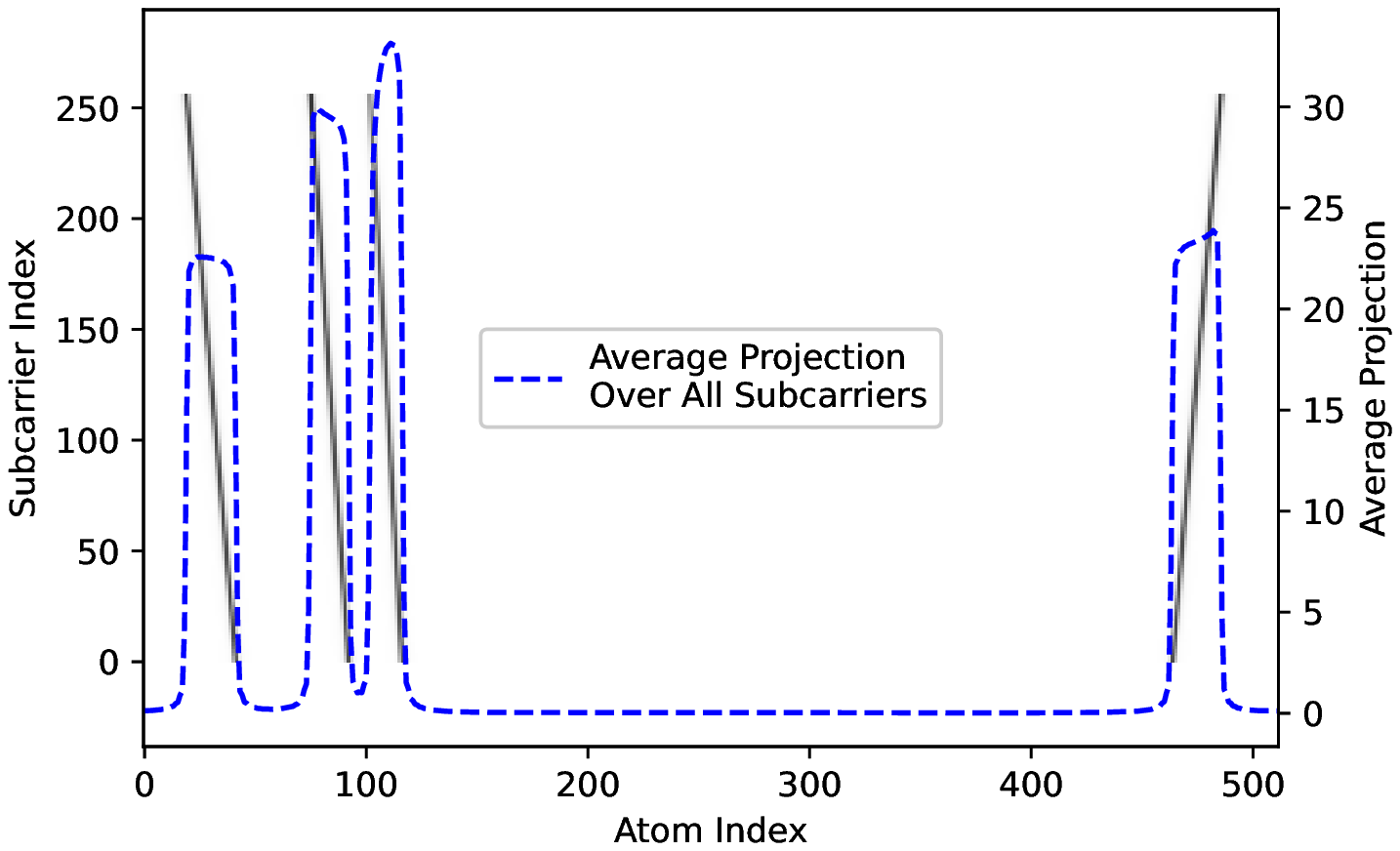}}
	\subfigure[With frequency-dependent $\widetilde{\bm{A}_k^{\text{ideal}}},\forall k$.]{
		\label{impact_of_measurement_matrix}
		\includegraphics[width=0.48\textwidth]{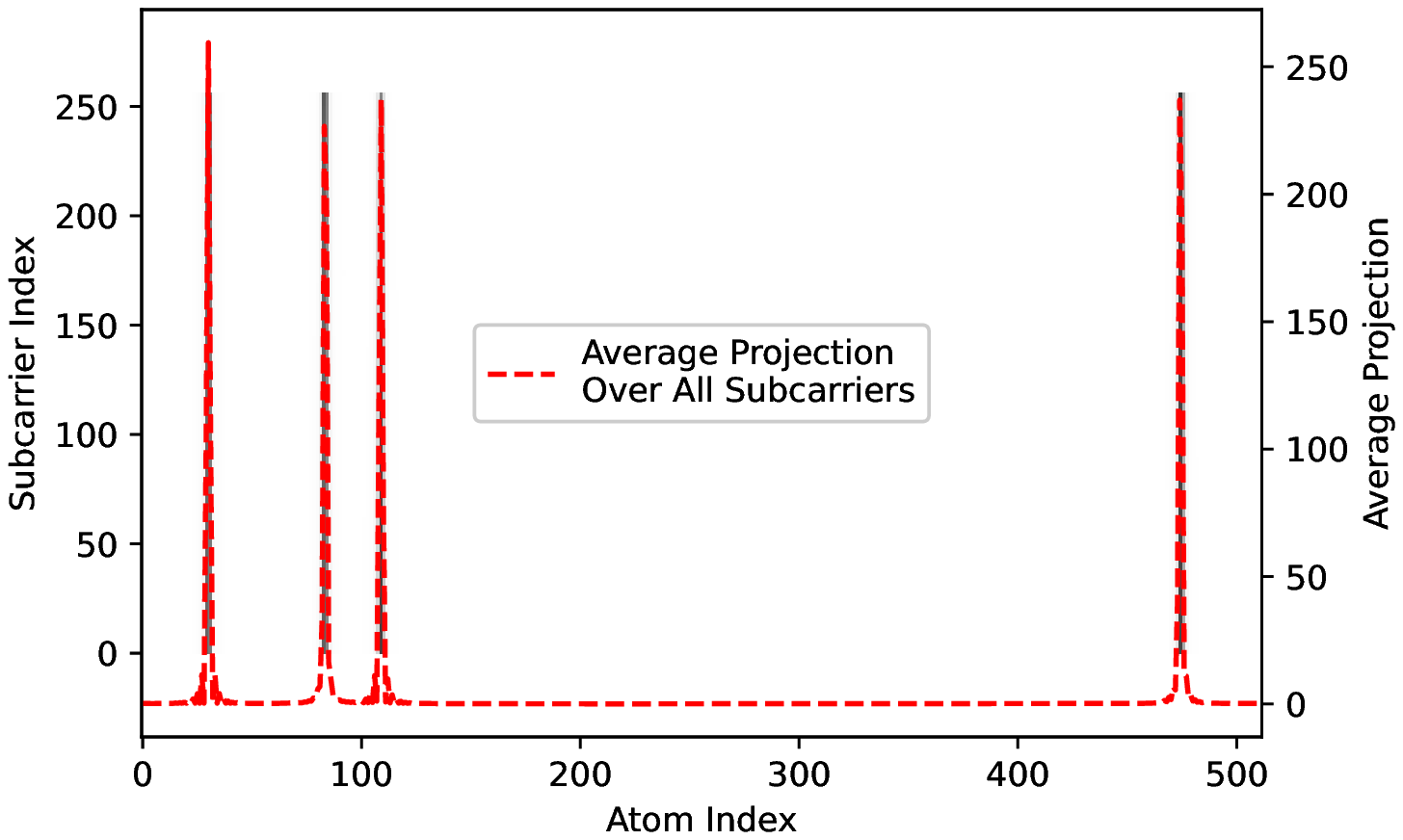}}
	\caption{The effectiveness of frequency-dependent measurement matrices in terms of support alignment. To make the figures clear, we set $\sigma_{\theta}=0^\circ,G=512,K=256$. Other parameters see the default setting in simulation.}
	\label{support_alignment}
\end{figure}

We remark that the factor $\frac{f_k}{f_c}$ in (\ref{grids}) plays the role of \emph{predistortion} such that the sparse supports of projections at all subcarriers would be aligned after experiencing the distortion of beam split. Fig. \ref{support_alignment} illustrates the effectiveness of the proposed frequency-dependent measurement matrices in terms of support alignment, where the grayscale images are summed projections of streams at different subcarriers and angles while the dashed curves are obtained by further averaging over all subcarriers. As we can see, with the frequency-independent measurement matrix, the sparse supports of projections at different subcarriers are not aligned. In contrast, with the proposed frequency-dependent measurement matrices, perfect common sparsity structure is indicated by the vertical straight lines on the grayscale image and the sharp peaks of the dashed curve. 

We would like to mention that we have also tried to use broader beams\cite{gff_heuristic} as atoms to cover wider angular ranges of clusters and optimize the measurement matrices based on channel samples using dictionary learning\cite{dictionary_learning}. Nevertheless, no performance gain is observed since the energy distribution in the angular domain of cluster channels changes from sample to sample, thus cannot be dynamically matched by any fixed measurement matrices. As a result, although these designs have different sample-wise performance, their average performance on a large number of samples is similar. 

\subsection{Feasible Phase and Delay Codebooks}
\label{codebook_design}
So far, feasibility issues are not considered yet. In practice, a group of analog precoders at all subcarriers are feasible only when specific phase and delay parameters satisfying respective constraints can be found. Therefore, we need to further optimize a feasible phase codebook, $\widetilde{\bm{A}}$, and a feasible delay codebook $\widetilde{\bm{T}}$, that can be implemented by actual hardware, such that the corresponding feasible measurement matrices are as close to the ideal measurement matrices as possible. Mathematically, the optimization problem is formulated as follows:
\begin{alignat}{2}
	\mathcal{P}4:\quad \mathop{\mathrm{min}}\limits_{\widetilde{\bm{A}},\widetilde{\bm{T}}}  &  \quad \sum_{k=1}^{K}\norm{\widetilde{\bm{A}_k^{\text{ideal}}}-\widetilde{\bm{A}}\odot (\widetilde{\bm{T}_k}\otimes\bm{1}_M)}^2_F, \\
	\mbox{s.t.} &\quad|\widetilde{A}(n_T,g)|=1, \forall n_T,g, \\
	&\quad\widetilde{T}_k(n_{TTD},g)=e^{-j2\pi f_k\widetilde{T}(n_{TTD},g)},\forall n_{TTD},g,k, \\
	&\quad0\ll \widetilde{T}(n_{TTD},g) \ll t_{max},\forall n_{TTD},g.
	\label{P4}
\end{alignat}
	
As a special case of $\mathcal{P}1$ when $N_s=G$ and the digital precoders are ignored, most existing algorithms for $\mathcal{P}1$ can be used to solve $\mathcal{P}4$ with slight modifications, such as the AM algorithm proposed in \cite{Alt_Opt}. Nevertheless, we propose a simple yet effective approach for codebook design, which can achieve low matrix approximation error with low complexity. To simplify the solution process, the following two Lemmas will be used, which are proved in Appendices A and B.

\noindent \textbf{Lemma 1.} \emph{The impact of adding any bias to all subcarriers' frequencies can be compensated by modifying the phase shifts properly according to the delays of TTD lines}.

\noindent \textbf{Lemma 2.} \emph{Adding any bias to the delays of all TTD lines connected to an RF chain will not change the sum rate}.

Let us first consider $N_{TTD}=N_T$, in which case the ideal measurement matrices are exactly achievable and the matrix approximation error can be zero. Since phase shifts of phase shifters are frequency-independent, it is straightforward to use the same phase codebook as the narrowband scenario in (\ref{matrix_narrowband}). Denote the initial phase codebook as $\widetilde{\bm{A}'}$, we set $\widetilde{\bm{A}'}=\widetilde{\bm{A}}$. Then, TTD lines are exploited to compensate for frequency-dependent phase differences between $\widetilde{\bm{A}'}$ and $\widetilde{\bm{A}_{k}^{\text{ideal}}},\forall k$. Considering the $g$-th atom, $n_T$-th antenna, and $k$-th subcarrier, we should let
\begin{equation}
e^{-j2\pi(f_k-f_c)\widetilde{T'}(n_T,g)}=\widetilde{\bm{A}_{k}^{\text{ideal}}}(n_T,g)-\widetilde{\bm{A}'}(n_T,g)=e^{-j\pi(n_T-1)(f_k-f_c)\phi^g/f_c},
\label{pd_compensation}
\end{equation}
where a bias of $-f_c$ is added to all subcarriers' frequencies according to {\bfseries Lemma 1} and $\widetilde{T'}(n_T,g)$ denotes the unconstrained delay of the $n_T$-th TTD line. Solving (\ref{pd_compensation}), we get the initial delay codebook as
\begin{equation}
\widetilde{T'}(n_T,g) = \frac{(n_T-1)\phi^g}{2f_c},\forall n_T,g.
\label{TTD_solution}
\end{equation}

However, the delay in (\ref{TTD_solution}) is negative when $\phi^g<0$. To meet the non-negative constraint of delays, {\bfseries Lemma 2} is used to obtain the final delay codebook as
\begin{equation}
	\widetilde{T}(n_T,g) = 
	\begin{cases}
	\widetilde{T'}(n_T,g) + \beta, &  0\le g \le G/2, \\
	\widetilde{T'}(n_T,g), & G/2< g \le G,
	\end{cases}
	\label{TTD_solution_final}
\end{equation}
where $\beta=\frac{(N_T-1)}{2f_c}$ equals the maximal propagation delay of the electromagnetic wave across the antenna array\cite{fixed_TTD,DPP}. 

Next, we consider the practical case where $N_{TTD}\ll N_T$ and the ideal measurement matrices can only be realized approximately. To minimize the approximation error, the computed delay of the middle antenna of each subarray is shared by all antennas in the subarray such that
\begin{equation}
	\widetilde{T}(n_{TTD},g) = 
	\begin{cases}
	\frac{(n_{TTD}'-1)\phi^g}{2f_c}+\beta, &  0\le g \le G/2, \\
	\frac{(n_{TTD}'-1)\phi^g}{2f_c}, & G/2< g \le G.
	\end{cases}
	\label{limited_TTD_solution_final}
\end{equation}
where $n_{TTD}'=(2Mn_{TTD}-M)//2+1$ denotes the index of the $n_{TTD}$-th subarray's middle antenna in the entire antenna array. When $N_{TTD}=N_T,M=1$, (\ref{limited_TTD_solution_final}) reduces to (\ref{TTD_solution_final}).

Once the delay codebook is determined, the initial phase codebook needs to be modified accordingly based on {\bfseries Lemma 1} to compensate for the impact of the previously added bias to all subcarriers' frequencies. With slight notation abuse, we still use $\widetilde{\bm{A}}$ to denote the final phase codebook in wideband DPP, which can be expressed as 
\begin{equation}
	\widetilde{\bm{A}} = \widetilde{\bm{A}'}\odot(\triangle\widetilde{\bm{A}'}\otimes\bm{1}_M),
	\label{phase_post_process}
\end{equation}
where $\triangle\widetilde{A'}(n_{TTD},g)=e^{j2\pi f_c \widetilde{T}(n_{TTD},g)},\forall n_{TTD},g$. Eventually, the feasible measurement matrices can be computed by
\begin{equation}
\widetilde{\bm{A}_k}=\widetilde{\bm{A}}\odot(\widetilde{\bm{T}_k}\otimes\bm{1}_M),\forall k,
\end{equation}
where $\widetilde{T_k}(n_{TTD},g)=e^{-j2\pi f_k \widetilde{T}(n_{TTD},g)}$ is proportional to the true subcarrier frequency. Using $\widetilde{\bm{A}_k},\forall k$, $\mathcal{P}1$ can be converted to the following MMV-CS problem $\mathcal{P}7$, where the constraint (32) is based on the fact that the sparse supports at all subcarriers are basically aligned even with beam split.
\begin{alignat}{2}
	\mathcal{P}7:\quad \mathop{\mathrm{min}}\limits_{\widetilde{\bm{D}_k}}  &  \quad \sum_{k=1}^{K}\norm{\bm{F}_k-\widetilde{\bm{A}_k}\widetilde{\bm{D}_k}}^2_F, \\
	\mbox{s.t.} & \quad \norm{\sum_{k=1}^{K}\left(\text{diag}\left(\widetilde{\bm{D}_k}\widetilde{\bm{D}_k}^H\right)\right)}_0=N_{RF}, \\
	&\quad\norm{\widetilde{\bm{A}_k}\widetilde{\bm{D}_k}}^2_F=N_s,\forall k,
	\label{P7}
\end{alignat}

\subsection{DPP Algorithm}
Given the above measurement matrices and codebooks, we extend {\bfseries Algorithm 1} to {\bfseries Algorithm 2} to solve $\mathcal{P}7$, which is named as Extended SSP (E-SSP). Compared to SSP, E-SSP has similar basic process while several major differences are emphasized as follows. First, the projections need to be computed at all subcarriers, so both the number of measurements and the number of sparse vectors are increased by a factor of $K$. Second, the summed projections of streams for atom selection require further averaging over all subcarriers. Last but not least, finer digital precoders are computed given the analog precoders. The previously used LS operation in SSP is optimal only in terms of matrix decomposition. To achieve our ultimate goal of sum rate maximization, SVD is executed on the equivalent channel of each subcarrier. The normalization term, $\left(\bm{A}_k^H\bm{A}_k\right)^{-\frac{1}{2}}$, in lines $18-19$ can transfer the power constraint from the hybrid precoder to the digital precoder\cite{gff_heuristic}, thus facilitating the calculation of the water-filling based optimal power allocation vector, $\bm{p}_{eq,k}$, in line $19$. 
\begin{algorithm}[htbp]
	\caption{E-SSP Based Wideband DPP}
	\textbf{Input}: $\widetilde{\bm{A}},\widetilde{\bm{T}},\widetilde{\bm{A}_k},\bm{F}_k,\bm{H}_k, \forall k,N_{RF},G,K;$ 
	\begin{algorithmic}[1]
		\State Initialize $\bm{A}^{(0)},\bm{T}^{(0)},\bm{A}_k^{(0)},\forall k$ as empty matrices, initialize $\bm{F}_{\text{res},k}^{(0)}=\bm{F}_k,\forall k;$  
		\State \textbf{for} $i=1:1:N_{RF}$ \textbf{do}  
		\State \quad $\bm{\psi}^{(i)}=\bm{0}_{G};$
		\State \quad\textbf{for} $k=1:1:K$ \textbf{do}
		\State \quad\quad $\bm{\Psi}_k^{(i)}=\widetilde{\bm{A}_k}^H\bm{F}_{\text{res},k}^{(i-1)};$  
		\State \quad\quad $\bm{\psi}_k^{(i)}=\text{diag}(\bm{\Psi}_k^{(i)}{\bm{\Psi}_k^{(i)}}^H);$
		\State \quad\quad $\bm{\psi}^{(i)}=\bm{\psi}^{(i)}+\bm{\psi}_k^{(i)};$
		\State \quad\textbf{end for}
		\State \quad $g^{(i)}=\mathop{\mathrm{argmax}}\limits_{j}\psi^{(i)}(j);$
		\State \quad $\bm{A}^{(i)}=[\bm{A}^{(i-1)}|\widetilde{\bm{A}}(:,g^{(i)})],\ \bm{T}^{(i)}=[\bm{T}^{(i-1)}|\widetilde{\bm{T}}(:,g^{(i)})], \ \bm{A}_k^{(i)}=[\bm{A}_k^{(i-1)}|\widetilde{\bm{A}_k}(:,g^{(i)})],\forall k;$ 
		\State \quad\textbf{for} $k=1:1:K$ \textbf{do}
		\State \quad\quad $\bm{D}_k^{(i)}={\bm{A}_k^{(i)}}^{\dagger}\bm{F}_k;$
		\State \quad\quad $\bm{F}_{\text{res},k}^{(i)}=\frac{\bm{F}_k-\bm{A}_k^{(i)}\bm{D}_k^{(i)}}{\norm{\bm{F}_k-\bm{A}_k^{(i)}\bm{D}_k^{(i)}}_F};$
		\State \quad\textbf{end for}
		\State \textbf{end for} 
		\State $\bm{A}=\bm{A}^{(N_{RF})},\ \bm{T}=\bm{T}^{(N_{RF})}, \ \bm{A}_k=\bm{A}_k^{(N_{RF})},\forall k;$
		\State \textbf{for} $k=1:1:K$ \textbf{do}
		\State \quad $\bm{H}_{eq,k}=\bm{H}_k\bm{A}_k\left(\bm{A}_k^H\bm{A}_k\right)^{-\frac{1}{2}}=\bm{U}_{eq,k}\bm{\Sigma}_{eq,k}\bm{V}_{eq,k}^H;$
		\State \quad $\bm{D}_k=\left(\bm{A}_k^H\bm{A}_k\right)^{-\frac{1}{2}}\bm{V}_{eq,k}(:,1:N_s)\text{diag}(\bm{p}_{eq,k});$
		\State \textbf{end for}
	\end{algorithmic}
	\textbf{Output}: $\bm{A},\bm{T},\bm{D}_k,\forall k;$
	\label{algorithm_essp}
\end{algorithm}

\section{Low-Complexity Extended SSP Based Wideband DPP}
The E-SSP algorithm proposed in the previous section can still have high complexity when the numbers of RF chains, subcarriers, and atoms are large. To enhance the practicality of the proposed approach, in this section, we reduce the complexity of E-SSP dramatically without sacrificing much performance using three efficient atom selection techniques.

\subsection{Non-Iterative Atom Selection}
\label{peak_finding}
The complexity of SSP is proportional to the number of iterations. In this subsection, we propose to discard the iterative process and select all atoms at once by exploiting the properties of projections in the angular domain.

In the original SOMP algorithm, the necessity of iteratively subtracting previously selected atoms' contributions comes from the correlation among different atoms. According to (\ref{ideal_matrices}), the atoms of the ideal measurement matrices\footnote{Although this technique is inspired by the properties of ideal measurement matrices, it also works with feasible measurement matrices thanks to the small matrix approximation error.} are in the form of ULA response vectors. It can be proved that two ULA response vectors corresponding to different angles are asymptotically orthogonal as the length of vectors tends to infinity\cite{massive_mimo2}. Since the number of antennas is relatively large in practical THz massive MIMO systems, the correlation among atoms would be relatively small. Therefore, it is reasonable to directly select the top-$N_{RF}$ atoms based on the initial projections without requiring multiple iterations. Nevertheless, when the angular distance between two atoms is very small, e.g., smaller than the angular Rayleigh resolution of the array\cite{massive_mimo3}, their mutual correlation is still non-negligible. To exclude atoms whose large projection comes from leaked power of neighbouring atoms, we further narrow down candidate atoms to only peaks on the initial projection curve\footnote{We have also tried to learn the mapping from the initial projections to the indexes of selected atoms with a deep neural network like \cite{DL_HBF1}. It turns out that these two methods have similar performance while finding peaks is much simpler and cheaper than network training.}, which can be found by various peak-finding algorithms.

\begin{figure}[!htb]
	\centering
	\includegraphics[width=0.65\linewidth]{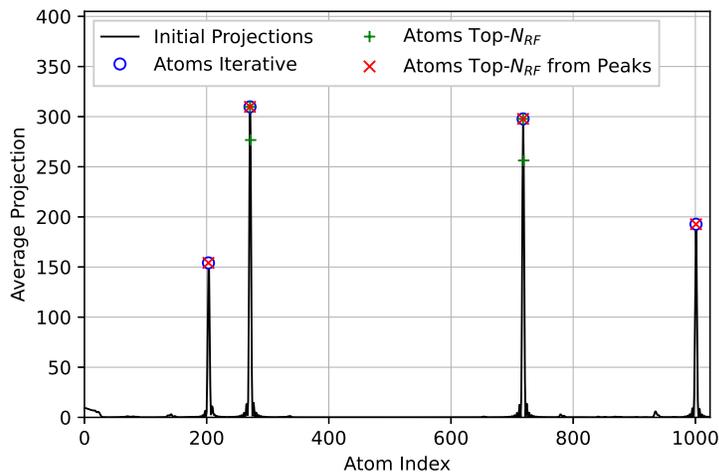}
	\caption{An exemplary initial projection curve and the selected atoms with different methods. To make the figure clear, we set $\sigma_{\theta}=0^\circ,G=1024$. Other parameters see the default setting in simulation.}
	\label{atom_selection}
\end{figure}

To promote intuitive understanding, one example of the initial projection curve and the atoms selected by different methods is given in Fig. \ref{atom_selection}. As we can see, directly selecting top-$N_{RF}$ atoms results in the selection of multiple nearby atoms with strong mutual correlation, as indicated by the pluses. In contrast, with the additional peak-finding based filtering, the same atoms as the original iterative method are selected, as indicated by the coincident crosses and circles. 

\subsection{Hierarchical Atom Selection and Refinement}
Although more atoms can improve the performance of E-SSP thanks to higher angular resolution, the complexity grows as well. To enjoy the benefit of high angular resolution with low complexity, in this subsection, we borrow the hierarchical beam sweeping idea from codebook-based precoding approaches\cite{subarray_codebook} to reduce the number of atoms that require projection computation.

Thanks to the continuity of the angles corresponding to atoms, nearby atoms have similar projections so that the projection curve is smooth in the angular domain. Therefore, the relative positions of peaks on the projection curve rarely change with $G$. For instance, if the $18$-th atom is a peak when $G=128$, it is very likely that there is also a peak around the $9$-th and the $36$-th atom when $G=64$ and $G=256$, respectively. Therefore, in the first stage, coarse-grained measurement matrices with only $G_c$ atoms ($G_c\ll G$), $\widetilde{\bm{A}^c_k},\forall k$, will be used to select $N_{RF}$ coarse-grained atoms, where the columns of each $\widetilde{\bm{A}^c_k}$ are evenly sampled from the corresponding fine-grained $\widetilde{\bm{A}_k}$ with $G$ atoms. Then, $2G_a$ fine-grained atoms will be added around each selected coarse-grained atom, to obtain new measurement matrices, $\widetilde{\bm{A}^f_k},\forall k$. In the second stage, atom selection procedures are executed once again using $\widetilde{\bm{A}^f_k},\forall k$, to determine the final fine-grained atoms. Using this two-stage hierarchical atom selection and refinement method, the number of atoms that require projection computation is at most ($G_c+2N_{RF}G_{a}$) while this number is $G$ if the fine-grained $\widetilde{\bm{A}_k},\forall k$, are directly used. Define the angular resolution reduction ratio as $\triangle G\triangleq G/G_c$, we will see that the performance degradation is marginal with a large $\triangle G$ and a small $G_a$ under typical system settings in simulation, thus reducing the complexity dramatically. 

\subsection{Partial Subcarrier Exploitation}
Since the number of subcarriers is usually large in practical wideband THz systems due to huge bandwidth, the complexity of E-SSP would be pretty high if projections at all subcarriers need to be computed. In this subsection, we propose to only exploit partial subcarriers' information for atom selection to reduce the complexity similar to \cite{partial_sc} which considers wideband channel estimation. 

When multiple sparse vectors have the perfect common sparsity structure, averaging the projections of their measurements can reduce the equivalent noise, thus improving performance. However, this benefit would saturate when the equivalent noise is low enough and no longer a performance bottleneck. For instance, the bottleneck could be the unrecoverable information loss in the imperfect fully-digital precoders in this case. Observing the CS model in (31), we find that the measurements, i.e., the fully-digital precoders at different subcarriers, are computed based on different subchannels. Therefore, the source of measurement noise is actually channel error, which can be caused by either imperfect estimation or feedback in practice. With lower channel error, fewer subcarriers are required to achieve a low enough equivalent noise level. In the extreme case where perfect CSI is available, only a single subcarrier is sufficient to achieve accurate atom selection with noiseless measurements. Nevertheless, the common sparsity structure among subcarriers is still imperfect due to finite angular resolution of atoms and the approximation error between the ideal and feasible measurement matrices. As as result, multiple measurements are still beneficial with perfect CSI while fewer measurements are required compared to the imperfect CSI case, as will be shown in simulation.

Denote the number of exploited subcarriers as $K'$, which can be evenly sampled from all subcarriers, the complexity of atom selection can be reduced by $\triangle K$ times, where $\triangle K \triangleq K/K'$ denotes the subcarrier reduction ratio. After the atoms are selected based on only $K'$ subcarriers' fully-digital precoders, the analog precoders at all $K$ subcarriers are determined while the digital precoders at all $K$ subcarriers can still be calculated in the same way. Notice that this technique is tailored for E-SSP and will lead to severe performance degradation if applied to other DPP algorithms, such as the AM algorithm proposed in \cite{Alt_Opt}. The reason is that without utilizing the highly structured codebooks, the optimized phases and delays that have good performance at a few sampled subcarriers are not necessarily good at the remaining subcarriers.

\subsection{Low-Complexity DPP Algorithm}
Overall, with the above three efficient atom selection techniques, the low-complexity extended SSP (LCE-SSP) algorithm for wideband DPP is given in {\bfseries Algorithm 3}. For compactness, several functions are used in the algorithm. Specifically, the $PeakFinder(\cdot,\bm{g}, N_{peak})$ function used in line $8$ and line $20$ returns the indexes of top-$N_{peak}$ peaks given index vector $\bm{g}$. In line $13$, the $IndexCleaner(\cdot,a,b)$ function cleans the index set by deleting duplicated elements and limiting the range of elements to $[a,b]$ while the $len(\cdot)$ function returns the length of the input vector.
\begin{algorithm}[!htb]
	\caption{LCE-SSP Based Wideband DPP}
	\textbf{Input}: $\widetilde{\bm{A}},\widetilde{\bm{T}},\widetilde{\bm{A}_k},\bm{F}_k,\bm{H}_k,\forall k,N_{RF},G,G_c,G_a,\triangle G,K',\triangle K;$
	\begin{algorithmic}[1]
		\State $\bm{g}^c=1:\triangle G:1+(G_c-1)\triangle G,\ \bm{\psi}^c=\bm{0}_{G_c};$
		\State \textbf{for} $k'=1:\triangle K:1+(K'-1)\triangle K$ \textbf{do}
		\State \quad $\widetilde{\bm{A}^c_{k'}}=\widetilde{\bm{A}_{k'}}(:,\bm{g}^c);$
		\State \quad $\bm{\Psi}^c_{k'}=\widetilde{\bm{A}^c_{k'}}^H\bm{F}_{k'};$ 
		\State \quad $\bm{\psi}^c_{k'}=\text{diag}(\bm{\Psi}^c_{k'}{\bm{\Psi}^c_{k'}}^H);$
		\State \quad $\bm{\psi}^c=\bm{\psi}^c+\bm{\psi}^c_{k'};$
		\State \textbf{end for}
		\State $\bm{g}^{c} = PeakFinder(\bm{\psi}^c,1:1:G_c,N_{RF});$
		\State $\bm{g}^f=$ empty vector;
		\State \textbf{for} $i=1:1:N_{RF}$ \textbf{do}
		\State \quad $\bm{g}^f=[\bm{g}^f|\triangle G\times g^{c}(i)-G_a:1:\triangle G\times g^{c}(i)+G_a];$
		\State \textbf{end for}
		\State $\bm{g}^f= IndexCleaner(\bm{g}^f,1,G),\ \bm{\psi}^f=\bm{0}_{len(\bm{g}^f)};$
		\State \textbf{for} $k'=1:\triangle K:1+(K'-1)\triangle K$ \textbf{do}
		\State \quad $\widetilde{\bm{A}^f_{k'}}=\widetilde{\bm{A}_{k'}}(:,\bm{g}^f);$
		\State \quad $\bm{\Psi}^f_{k'}=\widetilde{\bm{A}^f_{k'}}^H\bm{F}_{k'};$ 
		\State \quad $\bm{\psi}^f_{k'}=\text{diag}(\bm{\Psi}^f_{k'}{\bm{\Psi}^f_{k'}}^H);$
		\State \quad $\bm{\psi}^f=\bm{\psi}^f+\bm{\psi}^f_{k'};$
		\State \textbf{end for}
		\State $\bm{g} = PeakFinder(\bm{\psi}^f,\bm{g}^f,N_{RF});$
		\State $\bm{A}=\widetilde{\bm{A}}(:,\bm{g}),\ \bm{T}=\widetilde{\bm{T}}(:,\bm{g}), \ \bm{A}_k=\widetilde{\bm{A}_k}(:,\bm{g}),\forall k;$ 
		\State \textbf{for} $k=1:1:K$ \textbf{do}
		\State \quad $\bm{H}_{eq,k}=\bm{H}_k\bm{A}_k\left(\bm{A}_k^H\bm{A}_k\right)^{-\frac{1}{2}}=\bm{U}_{eq,k}\bm{\Sigma}_{eq,k}\bm{V}_{eq,k}^H;$
		\State \quad $\bm{D}_k=\left(\bm{A}_k^H\bm{A}_k\right)^{-\frac{1}{2}}\bm{V}_{eq,k}(:,1:N_s)\text{diag}(\bm{p}_{eq,k});$
		\State \textbf{end for}
	\end{algorithmic}
	\textbf{Output}: $\bm{A},\bm{T},\bm{D}_k,\forall k;$
	\label{algorithm_extended_ssp_low_complexity}
\end{algorithm}

\subsection{Baseline DPP Algorithms and Complexity Analysis} 
In this subsection, several major baseline DPP algorithms are introduced and the complexity of different DPP algorithms is analyzed and compared, as listed in Table \ref{complexity}. Only complexity-dominant operations are counted to make the $\mathcal{O}$ complexity expressions concise. To be fair, the digital precoders in all baseline algorithms are calculated in the same way as E-SSP and LCE-SSP, so this part of complexity is the same in all algorithms. The complexity difference mainly comes from different ways of obtaining the analog precoders. 

For all matrix decomposition based DPP algorithms, we first need $K$ matrix SVDs to obtain the fully-digital precoders. Then, in E-SSP demonstrated in {\bfseries Algorithm 2}, $3K$ matrix multiplications (line $5,12,13$) and $K$ Moore-Penrose matrix inverses (line $12$) are executed in each of the $N_{RF}$ iterations of atom selection. Finally, $2K$ matrix multiplications are needed to calculate digital precoders (line $18$). Notice that the complexity of codebook optimization is not included since it only occurs once offline for each system setting. 

Compared to E-SSP, the number of subcarriers involved in the calculation of fully-digital precoders in LCE-SSP reduces from $K$ to $K'$. Besides, the numbers of iterations, subcarriers, and atoms involved for atom selection in {\bfseries Algorithm 3} reduce from $N_{RF},K,G$ to $1,K'$, and at most $(G_c+2N_{RF}G_a)$, respectively.

In the AM algorithm proposed in \cite{Alt_Opt}, the phases, delays, and digital precoders are updated alternately when the other two are fixed in each iteration. The number of iterations is denoted by $N_{iter}$ and the update of delay parameters is based on the one-dimensional search among $S$ discrete delay grids. Besides, since AM's performance also depends on the initializer, the initialization algorithm's complexity denoted by $\mathcal{O}_{Init}$ also needs to be included in its overall complexity. 

In the heuristic DPP algorithm proposed in \cite{DPP}, fully-digital precoders are not needed since the analog precoders are computed based on angles of channel paths with very low complexity. Nevertheless, this algorithm cannot be directly applied to cluster channels where there are far more subpaths than RF chains. Simply using the mean angles of clusters (if assumed available) leads to poor performance. To include heuristic DPP as a meaningful baseline algorithm and enrich the comparison, we propose to find representative angles for it using the LCE-SSP algorithm with slight modifications. Specifically, we set $N_s$ to $N_R$ and replace the fully-digital precoders with channels in {\bfseries Algorithm 3}. Besides, ideal rather than feasible measurement matrices can be used since we are doing pure signal processing. In this way, $N_{RF}$ representative angles can be found such that the constructed pseudo path channels are close to the original cluster channels. We remark that such an extension of heuristic DPP to cluster channels is not trivial since the proposed algorithm is still used. The complexity of finding angles denoted by $\mathcal{O}_{FA}$ also needs to be included in the overall complexity of heuristic DPP.
\begin{table}[htb!]
	\begin{tabular}{cc}
		\toprule
		Algorithm & Complexity \\
		\midrule	
		E-SSP & $\mathcal{O}(KN_TN_R^2+KN_TN_{RF}(N_{RF}^2+N_{RF}N_s+N_sG+N_{RF}+N_R))$ \\
        LCE-SSP & $\mathcal{O}(K'N_TN_R^2+KN_TN_{RF}(2N_{RF}+N_R+N_s)+K'N_TN_s(G_c+2N_{RF}G_a))$\\
		AM & $\mathcal{O}_{Init}+\mathcal{O}\left(KN_TN_R^2+KN_TN_{RF}\left(N_{iter}(N_{RF}+N_s+S/M)+N_{RF}+N_R\right)\right)$ \\
		Heuristic DPP & $\mathcal{O}_{FA}+\mathcal{O}(KN_TN_{RF}(N_{RF}+N_R))$ \\ 
		\bottomrule
	\end{tabular}
	\centering
	\caption{Computational complexity of different DPP algorithms.}
	\label{complexity}
\end{table}

With typical values of system parameters and algorithm hyperparameters, the complexities of E-SSP and AM are comparable while the complexities of LCE-SSP and heuristic DPP are much lower. Intuitively, the average running time of LCE-SSP (several tens of milliseconds) can be dozens or even hundreds of times shorter than that of AM (several seconds) using the same programming language and CPU. 

We further remark that in FDD systems, the proposed E-SSP and LCE-SSP also have much lower feedback overhead\cite{ssp} than the AM algorithm. Thanks to the predetermined measurement matrices and codebooks, only $N_{RF}$ atom indexes are required to configure the hardware to construct the structured analog precoders at the BS, where each index needs $\mathrm{log}_2G$ bits to represent. In contrast, in the AM algorithm, each of the $N_{RF}N_{TTD}$ delay parameters needs $\mathrm{log}_2S$ bits to represent while the $N_{RF}N_T$ phase parameters are continuous (or quantized in practice) between $0$ and $2\pi$, thus resulting in a much higher feedback overhead.

\section{Simulation Results}
\label{simulation}
In this section, extensive numerical results\footnote{The source code for reproduction will be available at \href{https://github.com/EricGJB/Extended_SSP_Wideband_DPP}{$\mathrm{https://github.com/EricGJB/Extended\_SSP\_Wideband\_DPP}$} if the paper is accepted.} are provided to validate the superiority of the proposed approach in terms of performance, complexity, and robustness. Unless specified, the following system and channel parameters will be used as the default setting in simulation: $N_T=256$, $N_R=N_{RF}=N_s=4$, $N_{TTD}=16$, $f_c=100$ GHz, $f_s=10$ GHz, $K=128$, $N_c=4$, $N_p=10$, $\tau_{max}=20$ ns, $\sigma_{\tau_i}=1$ ns, $\sigma_{\theta^T_i}=\sigma_{\theta^R_i}=5^\circ$, $\forall i$, SNR $=10$ dB. The achievable data rate per subcarrier and the matrix approximation mean-squared error (MSE) defined as $\frac{1}{KN_TN_s}\sum_{k=1}^{K}\norm{\bm{F}_k-\bm{A}_k\bm{D}_k}_F^2$ are used as performance metrics. 

In E-SSP and LCE-SSP, hyperparameters with the best performance-complexity tradeoffs are searched. First, we gradually increase $G$ until the performance of E-SSP saturates. Then, using a large $G_a$, we gradually decrease $G_c$ until the performance degradation of LCE-SSP is non-negligible. Then, the same process is repeated to determine the smallest $G_a$ and $K'$. In this way, we get $G=1024,G_c=256,G_a=8,K'=4$. For heuristic DPP, mean angles of clusters can be used. Or, the representative angles can be found by CS algorithms such as the proposed LCE-SSP, as elaborated in Section IV.D. For the AM algorithm, we set $S=256$, $N_{iter}=20$, and increase $N_{iter}$ to $30$ in the multi-user case. All the simulation results are obtained by averaging over 200 Monte Carlo channel realizations. 

\begin{figure}[htb!] 
	\centering 
	\vspace{-0.35cm} 
	\subfigtopskip=2pt 
	\subfigbottomskip=1pt 
	\subfigcapskip=-3pt 
	\subfigure[Path channels with $\sigma_{\theta^T_i}=\sigma_{\theta^R_i}=0^\circ,\forall i$]{
		\label{impact_of_SNR_AS_0}
		\includegraphics[width=0.48\textwidth]{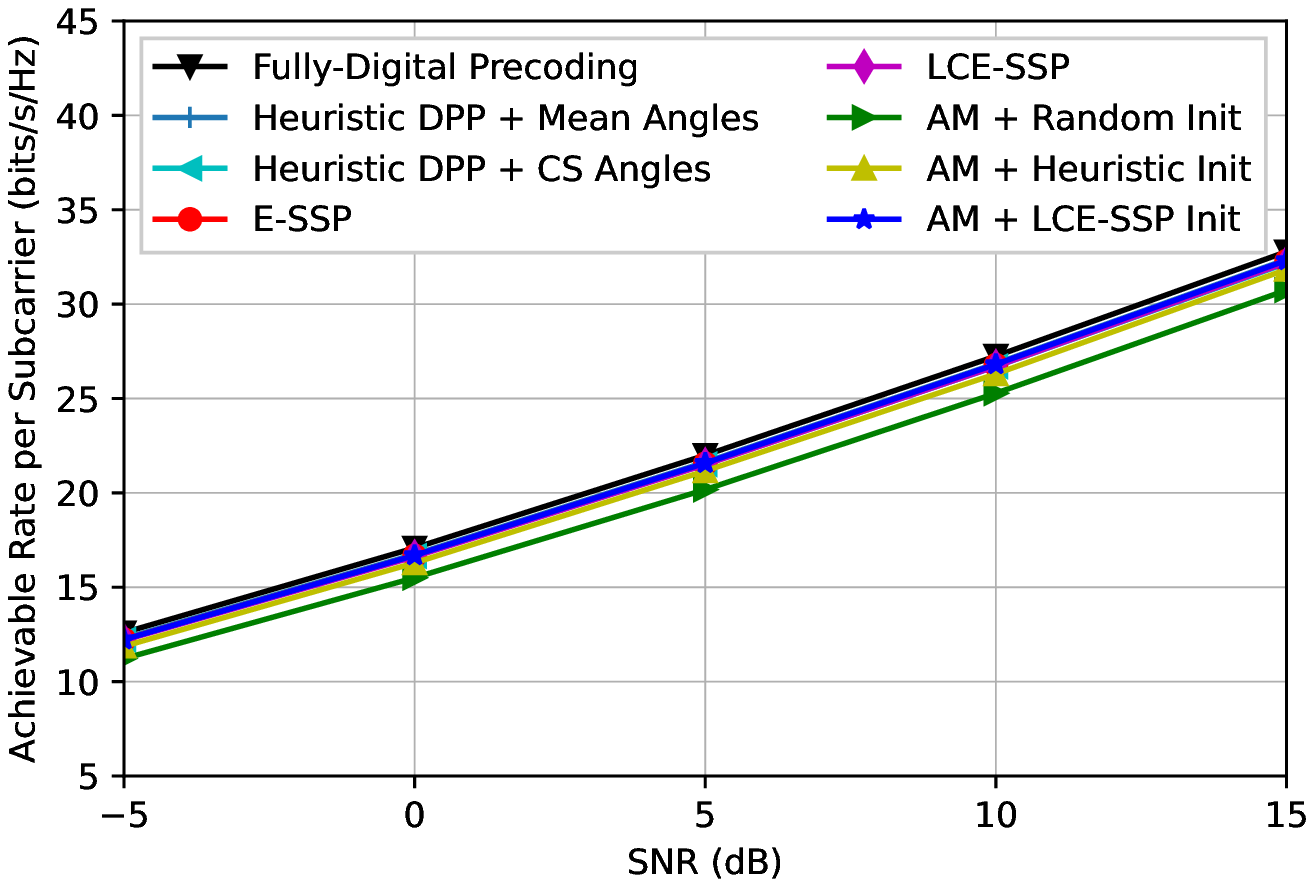}}
	\subfigure[Cluster channels with $\sigma_{\theta^T_i}=\sigma_{\theta^R_i}=5^\circ,\forall i$]{
		\label{impact_of_SNR_AS_5}
		\includegraphics[width=0.48\textwidth]{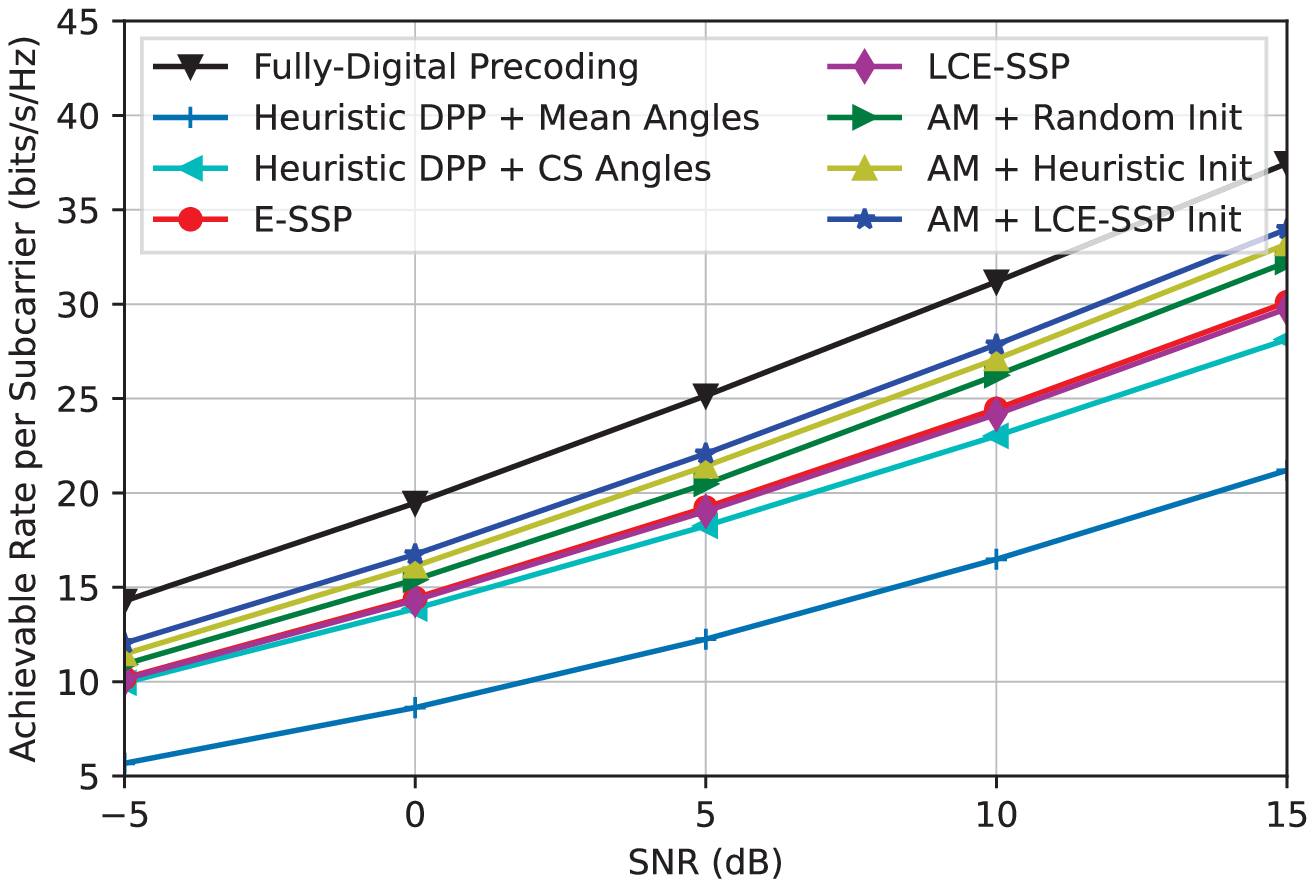}}
	\caption{Achievable rate versus SNR under different channel models.}
	\label{impact_of_SNR}
\end{figure}

Fig. \ref{impact_of_SNR} illustrates different DPP algorithms' achievable rate versus SNR under different channel models. First of all, we can see that LCE-SSP has almost the same performance as E-SSP in both path and cluster channels, demonstrating the effectiveness of the proposed low-complexity techniques. As shown in Fig. \ref{impact_of_SNR_AS_0}, in path channels, all DPP algorithms achieve close performance to fully-digital precoding in various SNR regimes thanks to the effective compensation for beam split. Further optimization using AM cannot improve the performance of initialization algorithms since they are already near-optimal. In contrast, according to Fig. \ref{impact_of_SNR_AS_5}, in cluster channels with a $5^\circ$ angular spread, the best DPP algorithm only achieves around $90\%$ of the performance upper bound due to more complicated channel distributions. Thanks to our extension, the heuristic DPP algorithm with angles found by the CS algorithm, i.e., LCE-SSP, performs much better than simply using mean angles of clusters. Nevertheless, it is still inferior to all three matrix approximation based algorithms, especially in high SNR regimes since they are not restricted to physical path angles but exploit the fully-digital precoders with richer information to guide the design of hybrid precoders. In AM, since the analog precoders do not have to be in specific forms, the generated beams would have larger degrees of freedom to flexibly cover each channel sample. Or, understood from another point of view, AM can be seen as a dictionary learning algorithm to find the optimal measurement matrices for E-SSP and LCE-SSP, but in a sample-wise manner. Although AM has high potentially achievable performance, its actual performance is also dependent on the quality of the initialization point. The role of a good initializer in AM is to construct initial analog precoders that are close to the optimal analog precoders to simplify the iterative optimization process and reduce the probability of falling into bad local optima. We can see from Fig. \ref{impact_of_SNR_AS_5} that LCE-SSP beats the heuristic DPP initializer and the random initializer, therefore can be used as a cheap yet effective initializer for AM in practical cluster THz channels.

\begin{figure}[!htb]
	\centering
	\includegraphics[width=0.48\linewidth]{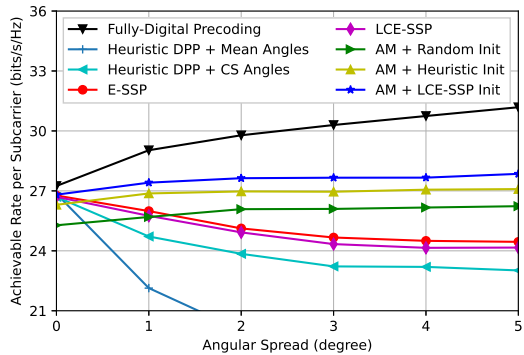}
	\caption{Impact of angular spread.}
	\label{impact_of_AS}
\end{figure}

In Fig. \ref{impact_of_AS}, the impact of angular spread is examined at a finer granularity. The performance of E-SSP and LCE-SSP decreases with angular spread since the fully-digital precoders for complicated channels with larger angular spreads are less likely to be well approximated by only a few atoms in the predetermined measurement matrices. Nevertheless, they are still indispensable to guarantee the high performance of AM as good initializers. Besides, when the angular spread is small, e.g., $\sigma_{\theta}<1^\circ$, LCE-SSP alone outperforms AM with the random initializer. The simultaneous good performance and low complexity makes LCE-SSP very appealing to such channel conditions. At last, without exploiting the fully-digital precoders, heuristic DPP is always inferior to LCE-SSP even with CS angles, either applied alone or as AM's initializer.

\begin{figure}[htb!]
	\centering
	\includegraphics[width=0.48\linewidth]{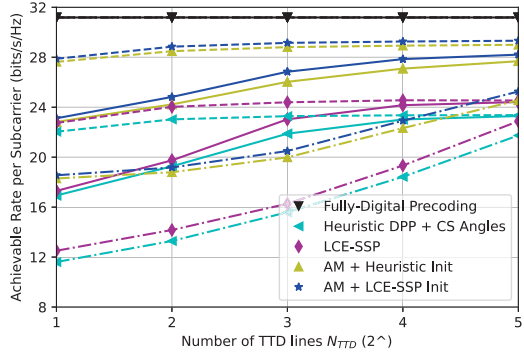}
	\caption{Impact of the remaining beam split level. The solid, dashed, and dash-dotted lines denote the fractional bandwidth being $0.1$, $0.025$, and $0.4$, respectively.}
	\label{impact_of_TTD_and_W}
\end{figure}

To promote understanding of the importance of beam split compensation, Fig. \ref{impact_of_TTD_and_W} illustrates the achievable rates with different remaining beam split levels, which are determined by both the number of TTD lines and the fractional bandwidth defined as $f_s/f_c$. On the one hand, with larger fractional bandwidth, the initial beam split level is higher, as indicated by different heights of the starting points of the curves. On the other hand, with more TTD lines, beam split can be compensated better, as indicated by the upward trend of the curves. The performance will not improve further when beam split is compensated thoroughly while the minimal number of TTD lines required to achieve this is proportional to the fractional bandwidth\cite{gff_heuristic}. For instance, the performance of LCE-SSP saturates with $32$ TTD lines when the fractional bandwidth is $0.1$ while only $8$ TTD lines are required when the fractional bandwidth is $0.025$. In contrast, the curve is still rising at $N_{TTD}=32$ when the fractional bandwidth is $0.4$.

\begin{figure}[htb!] 
	\centering 
	\vspace{-0.35cm} 
	\subfigtopskip=2pt 
	\subfigbottomskip=1pt 
	\subfigcapskip=-3pt 
	\subfigure[Objective convergence process.]{
		\label{objective_convergence}
		\includegraphics[width=0.48\textwidth]{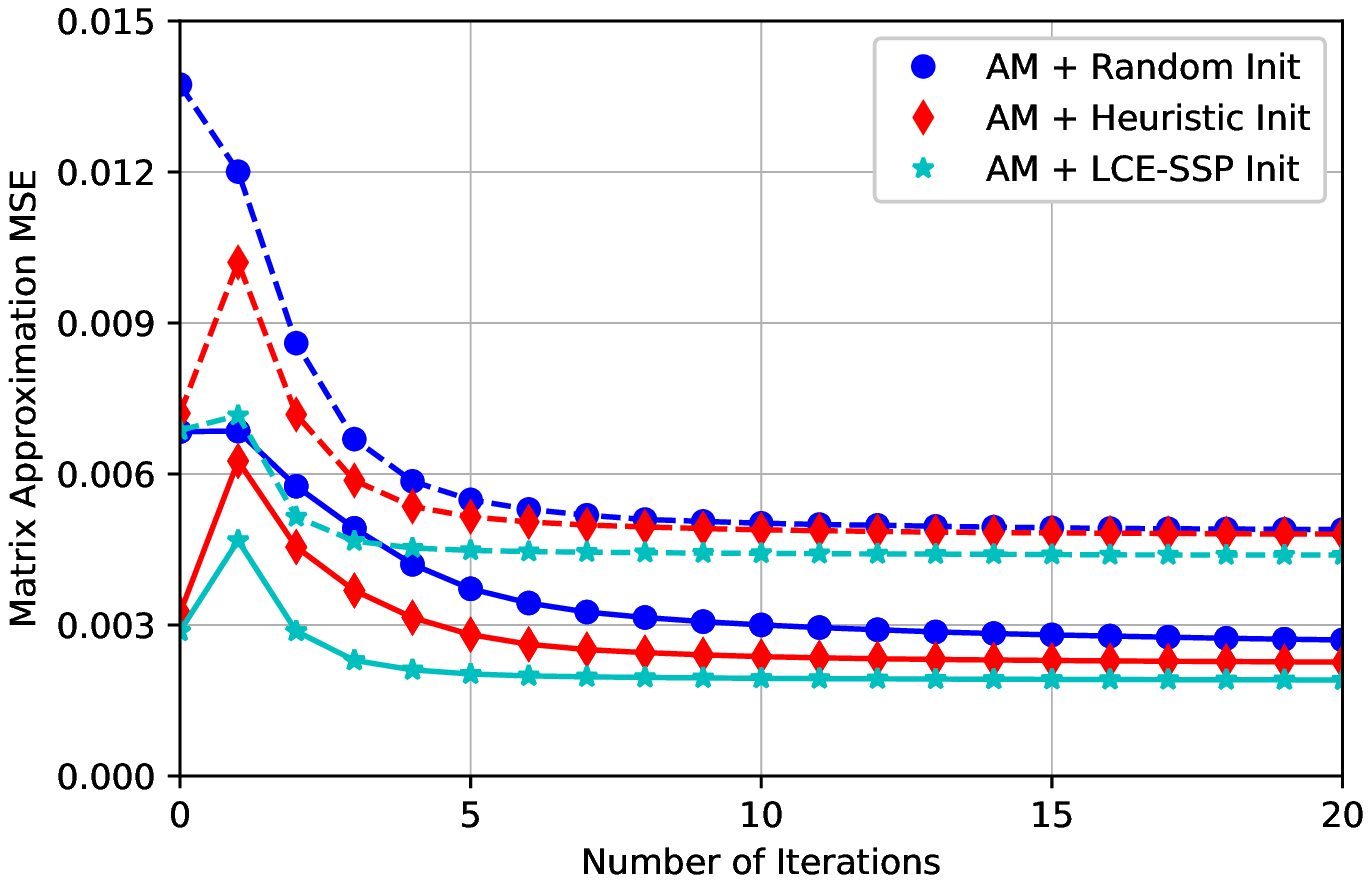}}
	\subfigure[Impact of early exit threshold.]{
		\label{impact_of_epsilon}
		\includegraphics[width=0.48\textwidth]{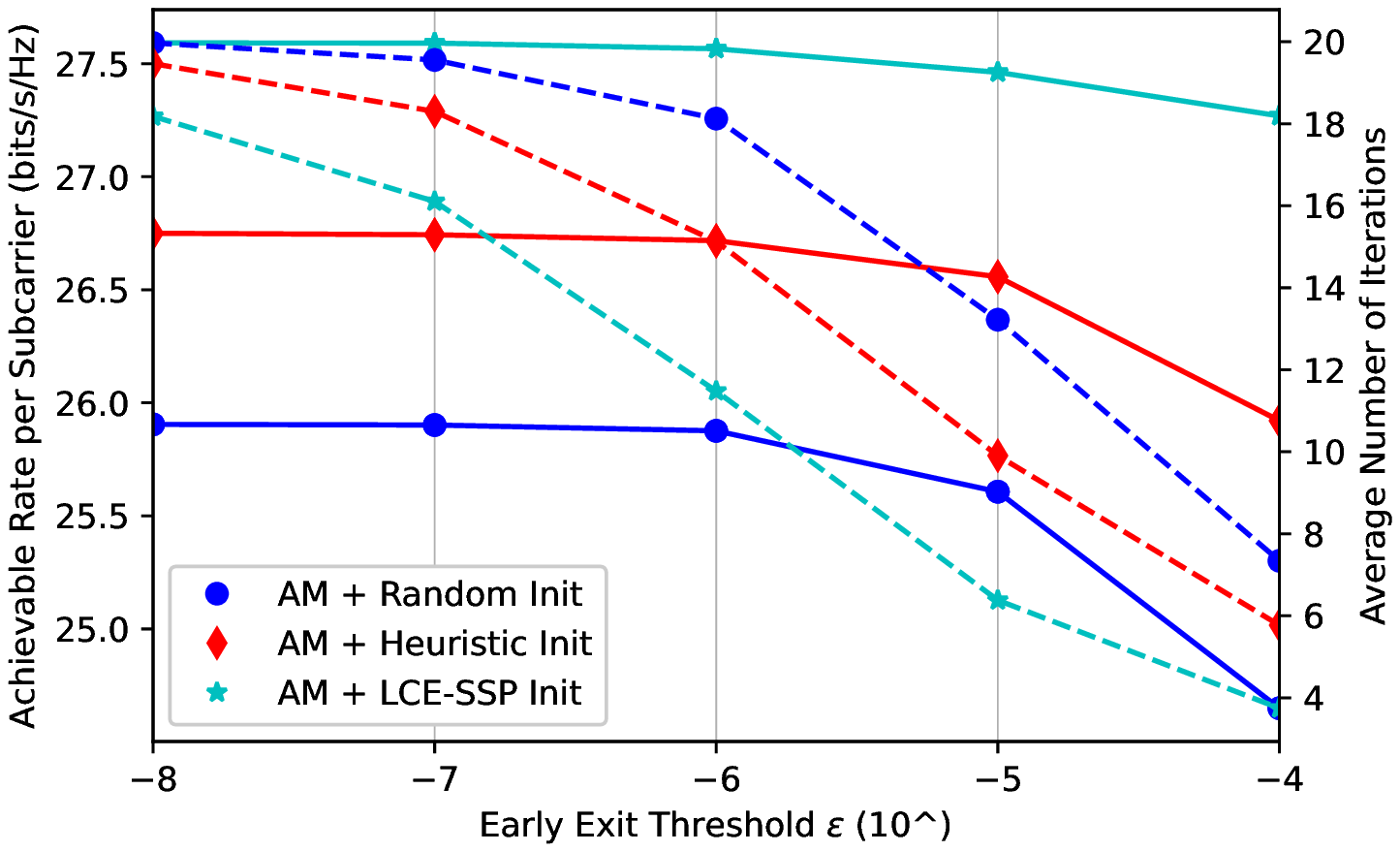}}
	\caption{Impacts of initialization methods on the convergence speed and performance of AM. In subfigure (a), the solid curves and the dashed curves denote $N_T=128,N_{TTD}=8,N_R=N_{RF}=N_s=2$ and $N_T=256,N_{TTD}=16,N_R=N_{RF}=N_s=4$, respectively. In subfigure (b), the solid curves and the dashed curves denote convergence performance and speed, respectively.}
	\label{impact_on_speed}
\end{figure}

Next, we show the impact of initializers on AM's convergence performance and speed in Fig. \ref{impact_on_speed}. According to Fig. \ref{objective_convergence}, both the convergence performance and speed of AM initialized by LCE-SSP are better than that initialized by the other two initializers in two different system scales, indicating its superiority. Same conclusions can be drawn from Fig. \ref{impact_of_epsilon}, where the early exit strategy is used to accelerate convergence\footnote{In many iterative algorithms, instead of running a fixed number of iterations, it is a common practice to stop the algorithm early when the decrease of objective between two iterations is small enough.}. Define the exit threshold of matrix approximation MSE in AM as $\epsilon$, we can see that $\epsilon=10^{-6}$ is a good choice to decrease the number of iterations without sacrificing performance. When $\epsilon=10^{-6}$, the average number of iterations of AM with the LCE-SSP initializer ($11$) is smaller than that with the heuristic DPP initializer ($15$) or with the random initializer ($18$). 

\begin{figure}[htb!] 
	\centering 
	\vspace{-0.35cm} 
	\subfigtopskip=2pt 
	\subfigbottomskip=1pt 
	\subfigcapskip=-3pt 
	\subfigure[Impact of channel error on achievable rate.]{
		\label{impact_of_CE_error1}
		\includegraphics[width=0.48\textwidth]{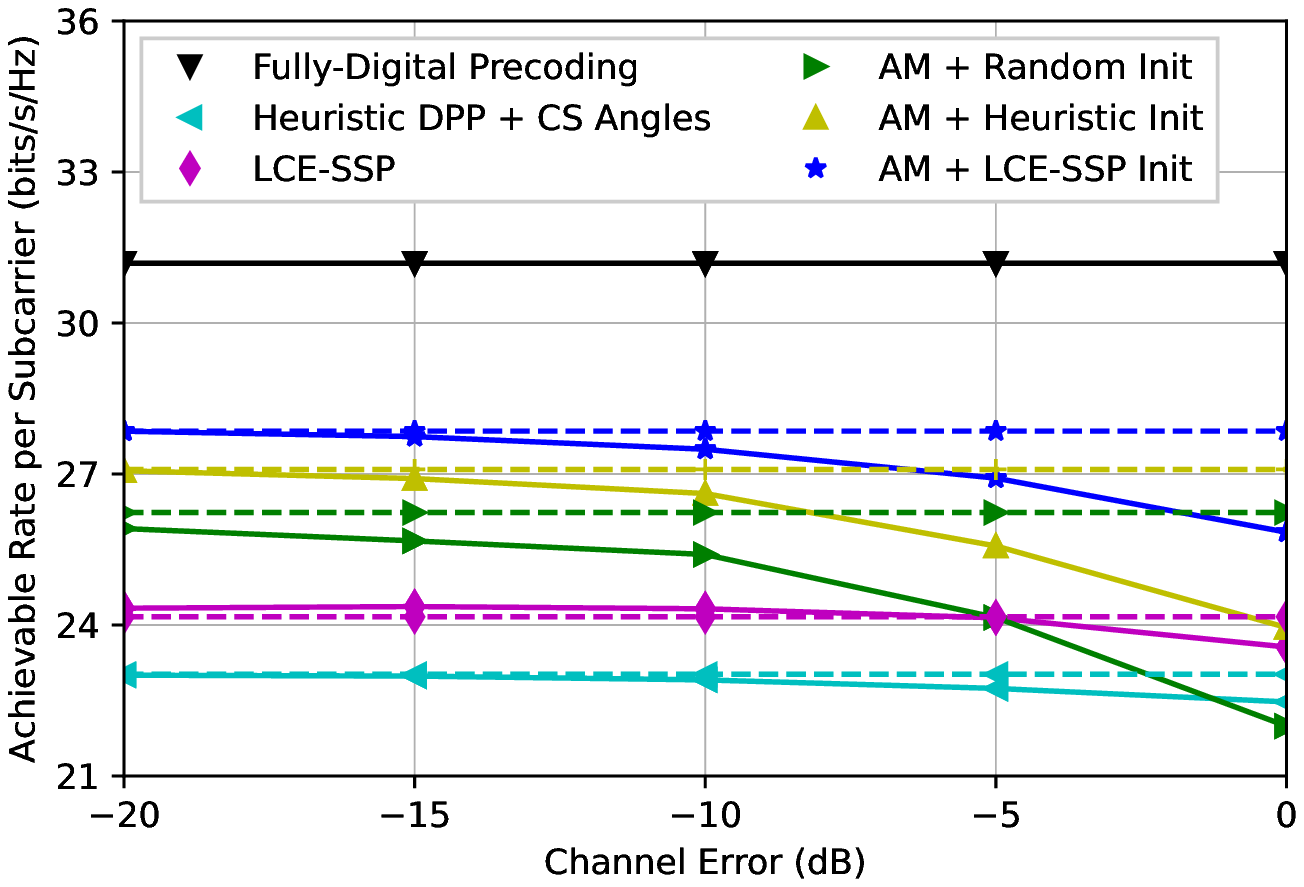}}
	\subfigure[Impact of channel error on $K'$ selection.]{
		\label{impact_of_CE_error2}
		\includegraphics[width=0.48\textwidth]{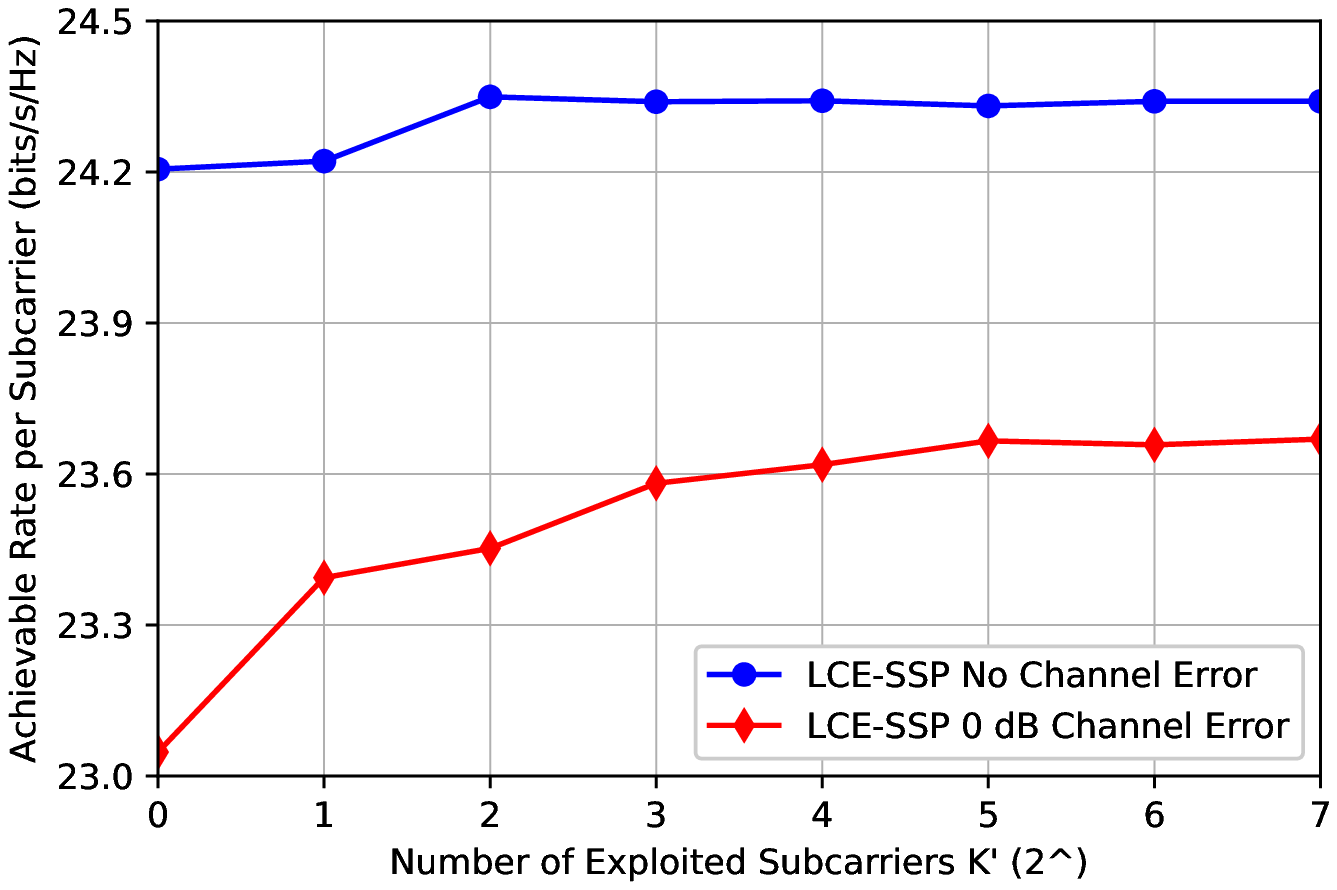}}
	\caption{Impacts of channel error on achievable rate and the selection of $K'$ in LCE-SSP.}
	\label{impact_of_CE_error}
\end{figure}

So far we have been assuming perfect CSI is available. However, channel estimation or feedback error is inevitable in practice. So, we investigate the impact of channel error, which is defined as $\mathbb{E}\left(\frac{1}{K}\sum_{k=1}^K\frac{\norm{\hat{\bm{H}}_k-\bm{H}_k}_F^2}{\norm{\bm{H}_k}_F^2}\right)$ with $\hat{\bm{H}}_k$ denoting the $k$-th estimated subchannel, in Fig. \ref{impact_of_CE_error}. From Fig. \ref{impact_of_CE_error1}, the performance degradation speed of heuristic DPP and LCE-SSP is slower than AM as channel error increases thanks to the restriction on analog precoders' structures. Specifically, the performance of AM starts to drop quickly when channel error is larger than $-15$ dB while the performance of the other two algorithms keeps almost unchanged until the channel error is larger than $-5$ dB. When the channel error is $0$ dB, LCE-SSP has similar performance as AM with the heuristic DPP initializer. Such strong robustness to imperfect CSI makes LCE-SSP appealing in challenging scenarios with large noises and limited resources during channel estimation and feedback. Initialized by LCE-SSP, AM still has the best performance thanks to the robust initialization point. As elaborated in Section IV.C, the selection of $K'$ depends on the channel error level. From Fig. \ref{impact_of_CE_error2}, we can observe that $4$ subcarriers are enough to achieve the saturated performance with perfect CSI, while this number increases to $32$ when the channel error is $0$ dB. With a proper selection of $K'$, the complexity of LCE-SSP is reduced dramatically without performance degradation compared to E-SSP.

\begin{figure}[htb!] 
	\centering 
	\vspace{-0.35cm} 
	\subfigtopskip=2pt 
	\subfigbottomskip=1pt 
	\subfigcapskip=-3pt 
	\subfigure[$N_U=4,N_{RF}=N_U$]{
		\label{impact_of_SNR_multi_user2}
		\includegraphics[width=0.48\textwidth]{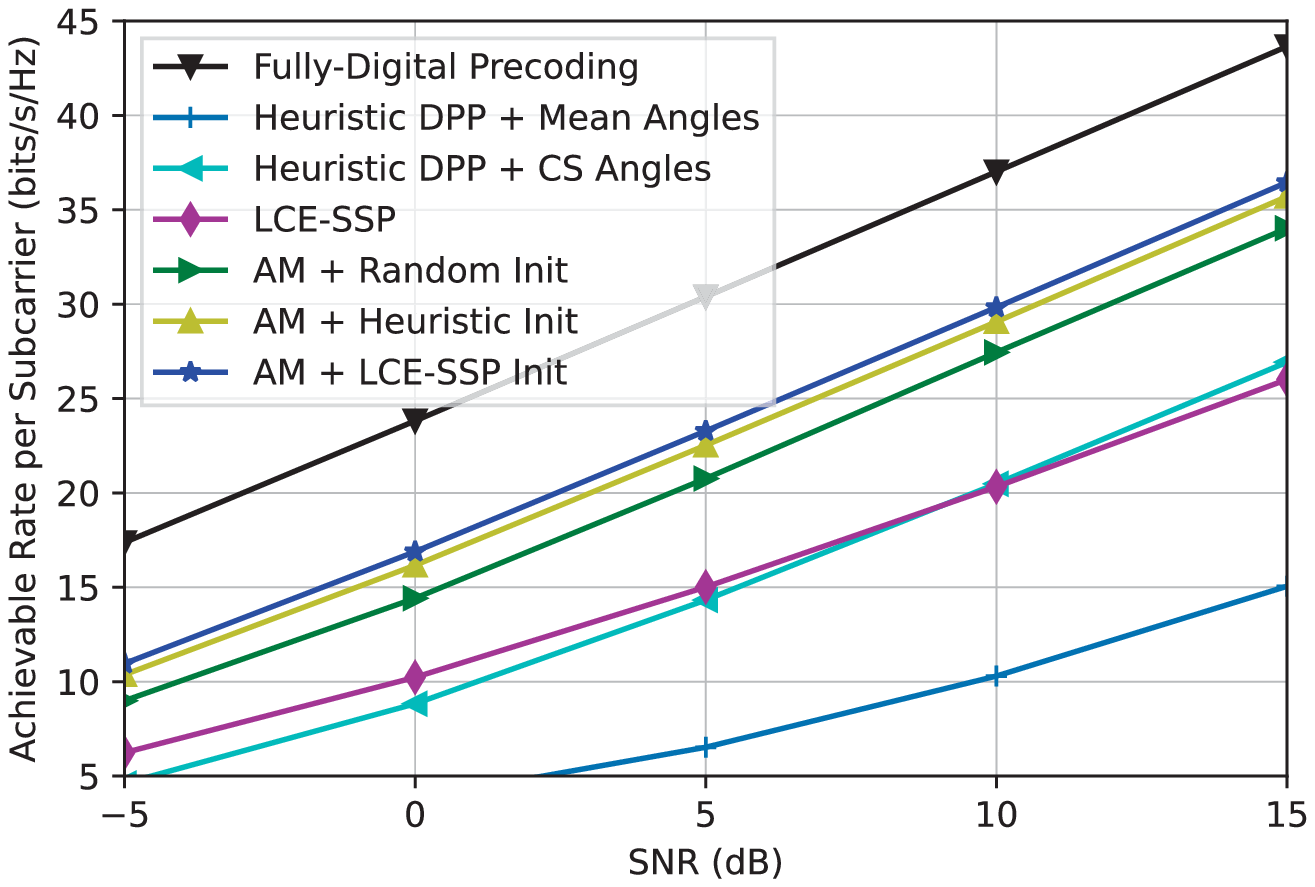}}
	\subfigure[$N_U=4,N_{RF}=N_UN_c$]{
		\label{impact_of_SNR_multi_user}
		\includegraphics[width=0.48\textwidth]{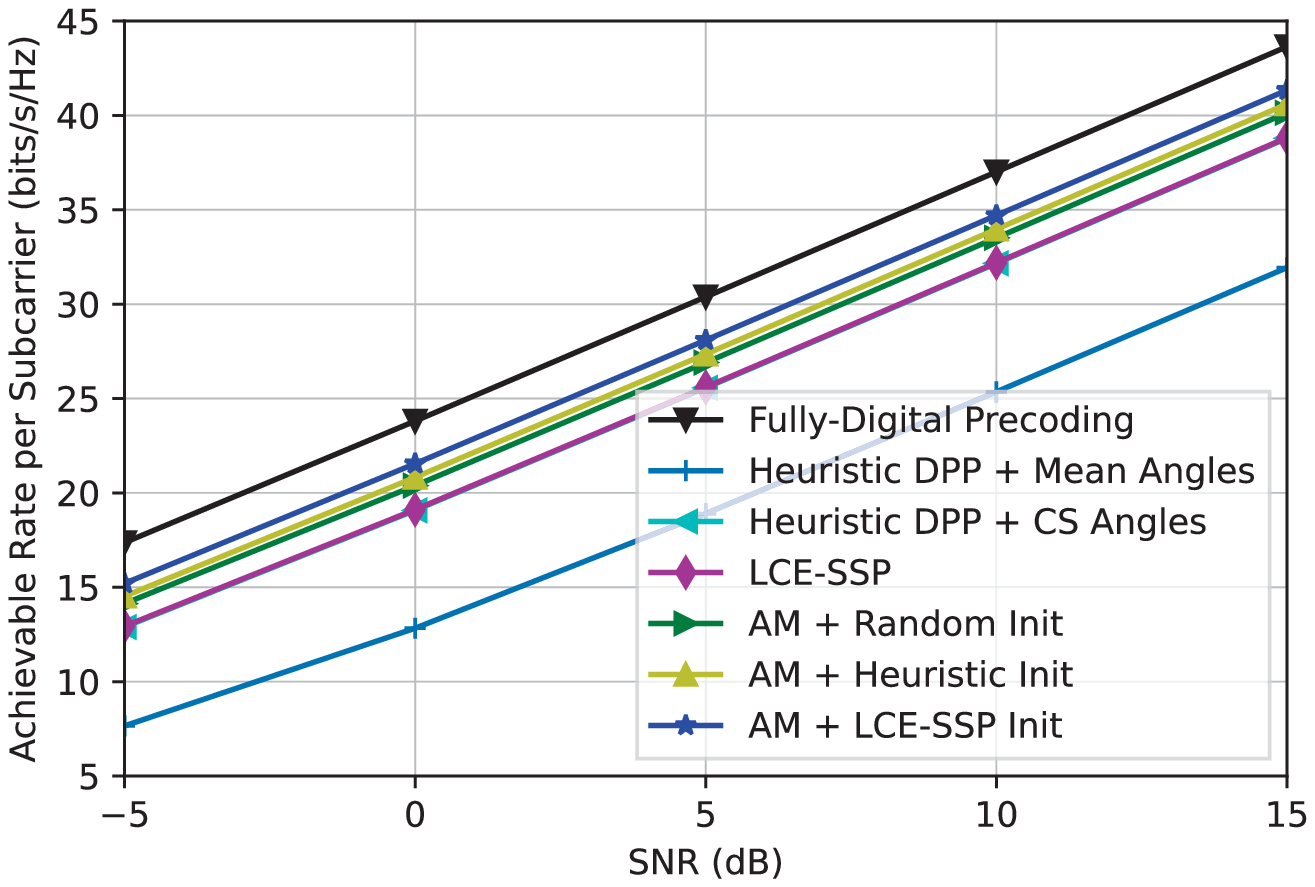}}
	\caption{Achievable rate versus SNR with different RF chain configurations.}
	\label{impact_of_SNR_MUE}
\end{figure}

The extension from the single-user scenario to the multi-user scenario is straightforward in matrix decomposition based DPP algorithms, as long as the fully-digital precoders are available\cite{Alt_Opt}. For the calculation of the fully-digital precoders and the digital precoders at the end, we perform zero-forcing (ZF) with water-filling based power allocation on the original channels and the equivalent channels, respectively. For heuristic DPP, $N_{RF}/N_U$ representative angles are contributed by each user's channels to make up all $N_{RF}$ angles, where $N_U$ denotes the number of users. We consider each user has a single antenna and one data stream as in \cite{Alt_Opt}. Two RF chain configurations are investigated, where $N_{RF}$ is set to $N_U$ to guarantee that each user is served in the first configuration and set to $N_UN_c$ to thoroughly exploit the spatial multiplexing gain in the second configuration. The performance when $N_U=4$ is illustrated in Fig. \ref{impact_of_SNR_MUE}. We can see that among all candidate initializers, LCE-SSP always leads to the best performance of AM. Specifically, it achieves $84\%$ performance of fully-digital precoding in the first configuration and this ratio increases to $95\%$ in the second configuration thanks to sufficient RF chains. Apart from better performance, the LCE-SSP initializer also leads to faster convergence of AM. For instance, when $N_{RF}=N_U,\epsilon=10^{-6}$, and SNR $=10$ dB, the average number of iterations with the LCE-SSP initializer is about half of that with the random initializer, indicating twice the speedup. In practice, the proposed LCE-SSP can be flexibly applied in different modes according to different requirements and RF chain configurations. When high performance is required and RF chains are limited, LCE-SSP can serve as a cheap yet effective initializer of AM. When low complexity is required and RF chains are sufficient, LCE-SSP alone can already achieve satisfactory performance. We also notice that the advantage of LCE-SSP over heuristic DPP is smaller than the single-user scenario since ZF is not the optimal fully-digital precoder in the multi-user scenario. Nevertheless, the good performance of heuristic DPP is still based on the representative angles found by LCE-SSP.

\section{Conclusion}
In this paper, we have proposed an extended SSP based DPP algorithm for THz massive MIMO systems. By exploiting the precoder structures, the matrix decomposition problem is converted to an MMV-CS problem, which can be readily solved by the slightly modified SOMP algorithm. To compensate for beam split, ideal frequency-dependent measurement matrices are designed for the alignment of sparse supports at all subcarriers, which can be approximately realized by optimized feasible phase and delay codebooks. To further reduce complexity, several efficient atom selection techniques are developed without sacrificing much performance. According to simulation results, by applied alone or in combination with existing DPP approaches, the proposed approach demonstrates superiority in performance, complexity, and robustness over existing approaches, thus is promising in future 6G THz massive MIMO systems.

\appendix
\subsection{Proof of {\bfseries Lemma 1}}
Consider a TTD line connected to an RF chain has delay $t$, and the $m$-th phase shifter connected to it has phase shift $e^{j\theta_m},\forall m$, then the superimposed phase of the TTD line and the $m$-th phase shifter at the $k$-th subcarrier is $e^{j(\theta_m-2\pi f_k t)},\forall m,k$. If we add a bias of $-\triangle f$ to all subcarriers' frequencies, the TTD line's phase at the $k$-th subcarrier will change from $e^{-j2\pi f_k t}$ to $e^{-j2\pi (f_k-\triangle f) t}$. To keep the superimposed phases unchanged, we need to modify the phase shifters' phase shifts to $e^{j(\theta_m-2\pi \triangle ft)},\forall m$. The reasoning process holds for any valid $\triangle f$, thus proving {\bfseries Lemma 1}.

\subsection{Proof of {\bfseries Lemma 2}}
Let us first consider a specific subcarrier and discard the subscript $k$. Denote the $N_{RF}\times 1$ vector of delay biases added to all TTD lines connected to $N_{RF}$ RF chains as $\triangle \bm{t}$, and denote the original analog precoder and the analog precoder with delay biases added as $\bm{A}$ and $\bm{A}'$, respectively, we have $\bm{A}'=\bm{A}\bm{\Sigma}$ and $\bm{A}=\bm{A}'\bm{\Sigma}^{-1}$, where $\bm{\Sigma}=\text{diag}\left(e^{-j2\pi f\triangle \bm{t}}\right)$ is an invertible diagonal matrix. Based on the properties of Moore-Penrose inverse such that $\bm{AA}^\dagger \bm{A}=\bm{A}$ and $(\bm{AB})^\dagger=\bm{B}^\dagger\bm{A}^\dagger$, we have 
\begin{equation}
	\bm{A}\bm{A}^{\dagger}\bm{A} = \bm{A}(\bm{A}'\bm{\Sigma}^{-1})^\dagger\bm{A} = \bm{A}\bm{\Sigma A}'^\dagger\bm{A}=\bm{A}.
	\label{q1}
\end{equation}

Right-multiply both sides of (\ref{q1}) by matrix $\bm{\Sigma A}'^\dagger$, we have
\begin{equation}
	\bm{A}\bm{\Sigma A}'^\dagger(\bm{A}\bm{\Sigma})\bm{A}'^\dagger = (\bm{A}\bm{\Sigma})\bm{A}'^\dagger,
\end{equation}
\begin{equation}
	\bm{A}\bm{\Sigma}(\bm{A}'^\dagger\bm{A}'\bm{A}'^\dagger) = \bm{A}'\bm{A}'^\dagger,
\end{equation}
\begin{equation}
	\bm{A}(\bm{\Sigma}\bm{A}'^\dagger) = \bm{A}'\bm{A}'^\dagger,
	\label{q2}
\end{equation}
where the property $\bm{A}^\dagger\bm{A}\bm{A}^\dagger=\bm{A}^\dagger$, is used to obtain (\ref{q2}). Eventually, based on $\bm{\Sigma}\bm{A}'^\dagger=\bm{A}^\dagger$, we prove that $\bm{A}\bm{A}^\dagger=\bm{A}'\bm{A}'^\dagger$. Since $\bm{A}\bm{A}^\dagger=\bm{A}'\bm{A}'^\dagger$ holds for any valid $\bm{A}$, for the $k$-th subcarrier, we have $\bm{A}_k\bm{D}_k=\bm{A}_k(\bm{A}_k^\dagger \bm{F}_k)=(\bm{A}_k\bm{A}_k^\dagger) \bm{F}_k=({\bm{A}'_k}{{\bm{A}'_k}^\dagger})\bm{F}_k={\bm{A}'_k}({{\bm{A}'_k}^\dagger}{\bm{F}_k})={\bm{A}'_k}{\bm{D}'_k}$, i.e., $\bm{A}_k\bm{D}_k$ does not change with $\triangle \bm{t}$. Therefore, the sum rate $R = \sum_{k=1}^{K}\mathrm{log}_2\left(\left|\bm{I}_{N_R}+\frac{\rho}{N_s\sigma_n^2}\bm{H}_k\bm{A}_k\bm{D}_k\bm{D}^H_k\bm{A}^H_k\bm{H}^H_k\right|\right)$ also won't change. The reasoning process holds for any valid $\triangle \bm{t}$, thus proving {\bfseries Lemma 2}.

\end{document}